\documentclass[%
 onecolumn,
superscriptaddress,
 amsmath,amssymb,
 aps,
 pra,
]{revtex4-1}

\usepackage{graphicx}
\usepackage{dcolumn}
\usepackage{bm}

\newtheorem{theorem}{\bf Theorem}[section]

\usepackage{amsmath}
\usepackage{bm}
\usepackage{color}
\usepackage{comment}
\usepackage{mathrsfs}
\usepackage{textcomp}

\newcommand \bk{\color{black}}

\usepackage[colorlinks=true, citecolor=red, linkcolor=blue,urlcolor=blue ]{hyperref}

\begin{document}

\preprint{APS/123-QED}

\title{The $N$-flagella problem:\\ Elastohydrodynamic motility transition of multi-flagellated bacteria}

\author{
Kenta Ishimoto}
\email{ishimoto@ms.u-tokyo.ac.jp}
\affiliation{Wolfson Centre for Mathematical Biology, Mathematical Institute, University of Oxford, Oxford OX2 6GG, UK}
\affiliation{Graduate School of Mathematical Sciences, The University of Tokyo, Tokyo 153-8914, Japan}
\author{
Eric Lauga}
\email{e.lauga@damtp.cam.ac.uk}
\affiliation{Department of Applied Mathematics and Theoretical Physics, University of Cambridge, Cambridge
CB3 0WA, UK}

\begin{abstract}
Peritrichous bacteria such as \textit{Escherichia coli} swim in viscous fluids by forming a helical bundle of flagellar filaments. The filaments are spatially distributed around the cell body to which they are connected via a flexible hook. To understand how the swimming direction of the cell is determined, we theoretically investigate the elastohydrodynamic motility problem of a multi-flagellated bacterium. Specifically, we  consider a spherical cell body with a number $N$ of flagella which are initially symmetrically arranged in a plane in order to provide an equilibrium state. We analytically solve the linear stability problem and find that at most 6 modes can be unstable and that these correspond to the degrees of freedom for the rigid-body motion of the cell body.
Although there exists a  rotation-dominated mode that generates negligible  locomotion, we show that for the typical morphological parameters of bacteria the most unstable mode results in linear swimming in one direction accompanied by rotation around the same axis, as observed experimentally.  
\end{abstract}

\maketitle

\section{Introduction}

Bacteria, which constitute the largest domain of prokaryotic microorganisms, have survived for billions of years due to  their sophisticated structures \cite{lighthill1976, bray2000,phillips2012}. Many bacteria are motile  and various forms of bacterial motility have been discovered, including gliding, twitching and swarming \cite{youderian1998,desiel2001,kearns2010,bergweb}. Above all, the most common form is swimming  and both its hydrodynamics basis and the chemotactic behaviour of cells have been investigated for many decades \cite{purcell1977,berg2003,berg2004,tindall2012,lauga2016}. 

Bacterial swimming is achieved using propulsion from helical appendages, called {\it flagellar filaments}, attached to the cell body (typical diameter $\approx 1-2~$\textmu m). A flagellar filament is a slender polymer, made of a single protein called flagellin, which takes the form of a rigid helix (typical length $\approx 10~$\textmu m and diameter $\approx 40$~nm), driven in rotation by a   bacterial rotary motor located at the base end of the filament.  The numbers and positions of the flagellar filaments can vary greatly from cell to cell, but so-called  \textit{peritrichous} bacteria possess multiple flagella effectively randomly distributed on the cell surface. This group of bacteria  includes  well-studied organisms such as \textit{Bacillus subtilis}, \textit{Salmonella enterica}  and the most popular model bacterium, \textit{Escherchia coli} \cite{berg2004,guttenplan2013,lauga2016}.

The behaviour of a rotary motor is regulated by inter-cellular signalling proteins. When a peritrichous bacterium is swimming (so-called ``run''), the distributed flagellar filaments  rotate in the same direction and gather together in a helical bundle, generating essentially linear  propulsion. When at least one of the rotary motors counter-rotates, the bundle of flagellar filaments comes apart and the cell changes its orientation \cite{macnab1977} (``tumble''). The resulting well-studied run-and-tumble mechanism allows peritrichous bacteria to explore chemically their environment.

Due to the small size of  bacterial cells, the typical   Reynolds number around a swimming bacterium is $Re\approx 10^{-4}$ and as a result, the fluid flow obeys the incompressible Stokes equations, which are time-reversible. The ability of bacteria to reorient requires them to break the time-reversibility constraint (i.e.~the scallop theorem \cite{purcell1977}), which is enabled by a short flexible hook ($\approx 60$~nm in length) that connects the rotary motor  to the semi-rigid flagellar filament. The motor/hook/filament complex is known as a flagellum. 
The flexibility of the hook allows it to behave as a universal joint \cite{samatey2004}  and is essential for flagellar bundle \cite{brown2012}. During the swimming motion, the flexible hook can buckle, providing a rich spectrum of swimming behaviours \cite{shum2012, son2013, nguyen2017, nguyen2018, jabbarzadeh2018}.

These mechanical structures together with the randomly distributed rotary motors of a peritrichous bacterium  raise a fundamental question: In which direction does the cell move   for a given motor configuration? The dynamics of a cell resulting from the rotating propulsion of   multiple rotating objects (the flagellar filaments) could be referred to as {\it $N$-flagella problem} in reference to the classical $N$-vortex problem on the  dynamics of  point vortices \cite{newton2001}. 
The flagellar morphologies, including pitch size and the radius of the helical filament, fall onto one of a small number of polymorphic shapes, which have been characterised experimentally  \cite{spagnolie2011} while the flagellar length can vary greatly  within species and cell populations. Despite these variations, a coherent flagellar bundle  is expected to be  maintained generically by  long-range hydrodynamic interactions  \cite{watari2010,kanehl2014}. 

Recently, Riley \textit{et al.} \cite{riley2018}    demonstrated theoretically that the   swimming of peritrichous bacteria is enabled by an elastohydrodynamic instability. Modelling  a bacterium as   propelled by multiple rigid helical filaments spatially distributed around the cell body and  connected  to it by linear torque springs, they  showed   that the coupling between the flagellar propulsive force pushing the cell body and the hydrodynamic forces resulting from the swimming motion could result in an elastohydrodynamic instability of the hook and lead to bacterial swimming towards a preferred direction. This motility transition, demonstrated numerically, was also explained  by a theoretical model of a bacterium propelled by two rod-like flagella connected to the spherical cell body from  opposite sides \cite{riley2018}.  The critical value predicted by the linear stability analysis was in good agreement with the full computational simulations, suggesting   that the swimming direction of peritrichous bacteria might be set by the stability of an equilibrium distribution of  flagellar filaments.

The aim of the current paper is to formalise this physical result mathematically  and to derive rigorously   the  elastohydrodynamic motility transition theory for a cell with an arbitrary number of flagella of arbitrary shape using linear stability analysis.  In the theoretical model of Ref.~\cite{riley2018}, the rod-like flagella only generate forces, which the issue of  torque generation was not considered. Since flagellar propulsion is generated by the rotation of helical filaments, a  generation of torque is inevitable  and we include it in this  paper. The number of flagella is set to  any integer $N$ greater than 1  and we will consider equilibria and its stability for the spatial distribution of the  flagella.

The paper is organised as follows.  
 Section \ref{sec:model} is devoted to the theoretical formulation of the problem.  
In Section \ref{sec:model}A, we formulate the bacterial motility problem for a cell with $N$ flagella, considering both force and torque balance for the cell and each flagellum.  In Section \ref{sec:model}B, we proceed to simplify the problem by focusing on the case of identical flagellar filaments generating  axisymmetric propulsion. The latter  property  holds for a helical filament assuming that  the timescale of the bending and rotation of the flagella are well separated, which is verified in practice. The typical flagellar rotation frequencies are  $\approx 100$~Hz and sufficiently faster than the typical frequency of cell rotation, $\approx 10$~Hz, that we may   approximate the flagellar propulsion by its time-average. We then consider the case where the $N$ flagella are symmetrically distributed in a plane around the cell body  and we focus on its linear stability from the equilibrium configuration,  which is presented in In Section \ref{sec:model}C. The  following sections are devoted to discussions of the  results of the linear stability analysis.  We first neglect small chirality effects in order to simplify the system, and  
in Section~\ref{sec:N=2} we start our analysis with the case of $N=2$ flagella, which is found to be different from the general case with $N>2$ due to the symmetry of the flagellar distribution. The general case is then discussed in Section~\ref{sec:generalN}, where we start with the examples of $N=3$ and $N=4$ before deriving rigorously the general stability results.  Finally,  we reincorporate the effects of chirality in Section~\ref{Sec:chilarity}  where solve the full problem.

\section{Mathematical model of peritrichous bacterium}
\label{sec:model}

\begin{figure}[t]
\centering\includegraphics[width=12cm]{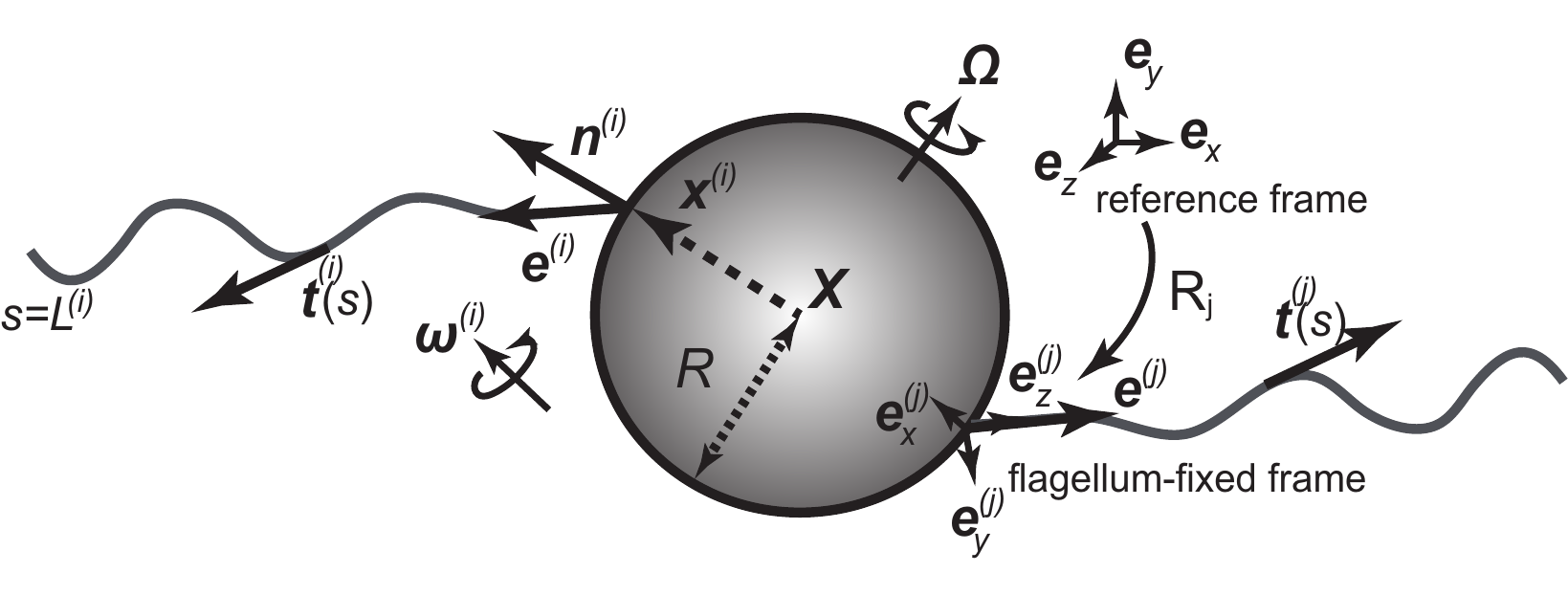}
\caption{Schematic representation of the model multi-flagellated bacterium considered in this paper (see text for details).}
\label{fig_model}
\end{figure}

\subsection{Equations of motion}
In this first section, we describe the force and torque balance equations, together with the torque balance of each elastic spring and hydrodynamic drag on each flagellar filament, in order to formulate a linear problem of $2N+6$ dimensions for  bacterial motility propelled by rotating  flagella.

We consider a swimming bacterium  located in a Newtonian fluid of constant dynamic viscosity $\mu$. The cell body is assumed to be a sphere of radius $R$. We denote the centre of the sphere by $\bm{X}$  and its orientation by a unit vector $\bm{e}$. The bacterium is assumed to possess $N$ rigid flagellar filaments connected to the cell body at their base. The direction for the axis of each filament  is measured by the unit   vectors $\bm{e}^{(i)}$ for $i$-th flagellum ($i=1, 2, \cdots, N$), as schematically illustrated in Fig.~\ref{fig_model}. We label each flagellar filament using the arc length, $s\in[0, L^{(i)}]$, measured from the flagellum-cell body connection (i.e.~the location of the motor), where $L^{(i)}$ is the length of the $i$-th flagellum, whose shape is determined by its tangent vector $\bm{t}^{(i)}(s)$.

\subsubsection{Force balance for a whole cell}

We first consider the force balance equations for the entire cell, which, in the absence of inertia, state that the sum of the hydrodynamic forces on the cell body and all flagella must add up to zero. 

Let $\bm{U}$ and $\bm{\Omega}$ be the linear and angular velocities of the cell, respectively. Neglecting   hydrodynamic interactions between the cell body and  the flagellar filaments,  the hydrodynamic drag on the cell body, $\bm{F}_{body}$,   is simply given from the  Stokes law, by $\bm{F}_{body}=C_D\bm{U}$, where $C_D=-6\pi\mu R$ and hereafter we non-dimensionalise the length scale using $R=1$. 

The hydrodynamic force on the $i$-th flagellum, $\bm{F}^{(i)}$, is obtained from the resistive-force theory of slender filaments, 
  which predicts that the hydrodynamic force on a small segment of the   flagellum, $d\bm{F}$, is linearly related to its local velocity relative to the background fluid, $\bm{u}^{(i)}$, as
\begin{equation}
d\bm{F}^{(i)}(s)=\left[C_t\bm{t}^{(i)}\bm{t}^{(i)}+C_n(\bm{1}-\bm{t}^{(i)}\bm{t}^{(i)})\right]\cdot\bm{u}^{(i)}(s)ds
\label{eq:m01},
\end{equation}
where $C_t$ and $C_n$ are the  negative drag coefficient constants whose values depend on the flagellar slenderness parameter\cite{gray1955,lauga2009}. 
Here, the local velocity $\bm{u}^{(i)}$ is the sum of the velocities of the cell body and the flagella. Introducing the angular velocity of each flagellum as $\bm{\omega}^{(i)}$ (Fig.~\ref{fig_model}), we have 
\begin{equation}
\bm{u}^{(i)}=\bm{U}+\bm{\Omega}\times(\bm{x}^{(i)}+\bm{\xi}^{(i)})+\bm{\omega}^{(i)}\times\bm{\xi}^{(i)}
\label{eq:m02},
\end{equation}
where 
\begin{equation}
\bm{\xi}^{(i)}(s)=\int_0^{s} \bm{t}^{(i)}(s')\,ds'
\label{eq:m03}
\end{equation}
is the position along the flagellar segment at arc length  $s$. From (\ref{eq:m02}), integrating over the entire flagellar filament leads to the hydrodynamic force on the $i$-th flagellum, with denoting the drag coefficient tensor in (\ref{eq:m01}) by $\mathsf{C}_i$, 
given by
\begin{equation}
\bm{F}^{(i)}=\left[ \int_0^{L^{(i)}} \mathsf{C}_i\,ds\right]\cdot\bm{U}
+\left[ \int_0^{L^{(i)}} \mathsf{C}_i\cdot \mathsf{A}_i\,ds\right]\cdot\bm{\Omega}
+\left[\int_0^{L^{(i)}} \mathsf{C}_i\cdot \tilde{\mathsf{A}}_i\,ds\right]\cdot\bm{\omega}^{(i)}
\label{eq:m04}.
\end{equation}
Here, we have introduced skew-symmetric matrices $\mathsf{A}_i$ and $\tilde{\mathsf{A}}_i$ whose components are given, respectively,  by 
\begin{equation}
\left[\mathsf{A}_i\right]_{pq}=\varepsilon_{pqr}(x^{(i)}_r+\xi^{(i)}_r)~~\textrm{and}~~\left[\tilde{\mathsf{A}}_i\right]_{pq}=\varepsilon_{pqr}\xi^{(i)}_r
\label{eq:m05},
\end{equation}
where $\varepsilon_{pqr}$ is the Levi-Civita symbol and the Einstein summation convention is used over the repeated indices ($p,q,r=1,2,3$). The final force balance equations  obtained by summing up the forces as
\begin{equation}
\bm{F}_{body}+\sum_{i=1}^N\bm{F}^{(i)}=\bm{0}.
\label{eq:06}
\end{equation}

\subsubsection{Torque balance for a whole cell}

Similarly to the above arguments, we now derive the expressions for the torque balance for an entire cell at the centre of the cell body. We again neglect hydrodynamic interactions between the cell and flagella  and model hydrodynamic drag on each flagellar filament at the level of resistive force theory (RFT).

The torque acting on the spherical cell body is given by $\bm{M}_{body}=C_R\bm{\Omega}$, where the resistance  coefficient is $C_R=-8\pi\mu R^3$. The hydrodynamic torque on a segment of a flagellum is given by $d\bm{M}^{(i)}=(\bm{x}^{(i)}+\bm{\xi}^{(i)})\times\,d\bm{F}^{(i)}$, which yields the total torque expression after an integration over the flagellum as
\begin{equation}
\bm{M}^{(i)}=\left[ \int_0^{L^{(i)}} \mathsf{A}^{T}_i\cdot\mathsf{C}_i\,ds\right]\cdot\bm{U}
+\left[ \int_0^{L^{(i)}}\mathsf{A}^{T}_i \cdot\mathsf{C}_i\cdot \mathsf{A}_i\,ds\right]\cdot\bm{\Omega}
+\left[\int_0^{L^{(i)}}\mathsf{A}^{T}_i \cdot\mathsf{C}_i\cdot \tilde{\mathsf{A}}_i\,ds\right]\cdot\bm{\omega}^{(i)}
\label{eq:10},
\end{equation}
where he superscript, $T$, indicates the transpose of a matrix.  Note that in order to obtain that expression we neglected the torque arising from local rotation of the flagellar filaments around their centreline, which is typically orders of magnitude smaller than (\ref{eq:10}) in the experimental limit where the radius of the helical centreline is  much larger than the thickness of the filament. 

The overall  torque balance equation for the whole cell is then written
\begin{equation}
\bm{M}_{body}+\sum_{i=1}^N\bm{M}^{(i)}=\bm{0}
\label{eq:11}.
\end{equation}

\subsubsection{Torque balance for each flagellum}

We  proceed to consider the torque balance relation for each flagellum, which experiences both hydrodynamic torque and an elastic spring restoring torque at the flagellum-cell body junction \cite{ishimoto2017,riley2018}. The hydrodynamic torque follows from   the discussion  above  and we now  consider the torque balance at the flagellum-cell body junction point. The torque acting on a segment of a flagellum is then given by $d\tilde{\bm{M}}=\bm{\xi}^{(i)}\times d\bm{F}^{(i)}$  and integrating over the flagellum yields the hydrodynamic torque,
\begin{equation}
\tilde{\bm{M}}^{(i)}=\left[ \int_0^{L^{(i)}} \tilde{\mathsf{A}}^{T}_i\cdot\mathsf{C}_i\,ds\right]\cdot\bm{U}
+\left[ \int_0^{L^{(i)}}\tilde{\mathsf{A}}^{T}_i \cdot\mathsf{C}_i\cdot \mathsf{A}_i\,ds\right]\cdot\bm{\Omega}
+\left[\int_0^{L^{(i)}}\tilde{\mathsf{A}}^{T}_i \cdot\mathsf{C}_i\cdot \tilde{\mathsf{A}}_i\,ds\right]\cdot\bm{\omega}^{(i)}
\label{eq:20}.
\end{equation}

As a model for the elastic hook, a linear spring torque is assumed to be present at the junction point. Let $\kappa^{(i)}>0$ be the spring constant  and let the relative angle difference from the initial orientation be denoted by $\theta^{(i)}$. The elastic torque on each flagellum can be then written as 
\begin{equation}
\bm{M}^{(i)}_{elast}=-\kappa^{(i)}\theta^{(i)}\bm{e}^{(i)}_{\perp},
\end{equation}
 where $\bm{e}^{(i)}_{\perp}$ is the unit vector   perpendicular both to the initial and current flagellar orientation vectors. In the later part of this manuscript, we will assume that the flagellar orientations initially coincide with the outward normal $\bm{n}^{(i)}$ (Fig.~\ref{fig_model}), but the formulation here does not necessarily assume this initial condition.

Hereafter, we assume that each flagellum is rotated at  a fixed rate  and thus the torque balance equation for each flagellum is obtained as the instantaneous balance
\begin{equation}
\mathsf{P}_i\cdot\tilde{\bm{M}}^{(i)}+\bm{M}^{(i)}_{elast}=\bm{0}
\label{eq:21a},
\end{equation} 
where we note that only the torque balance perpendicular to the vector $\bm{e}^{(i)}$ is considered, where $\mathsf{P}_i=(\mathsf{1}-\bm{e}^{(i)}\bm{e}^{(i)})$ is the projection onto the plane perpendicular to $\bm{e}^{(i)}$.

Note that if, alternatively,  one was to model  swimming as induced by motors rotating at fixed torque  \cite{shum2012,kanehl2014},  we would need an additional term in the torque balance equation, which would take the form $\tilde{\bm{M}}^{(i)}+\bm{M}^{(i)}_{elast}+\bm{M}^{(i)}_{motor}=\bm{0} \label{eq:21b}$, where now it is the moment $\bm{M}^{(i)}_{motor}$  which has a fixed value. In what follows, we focus   on the rotation-given problem, aiming to generalise the results predicted  by the  previous theoretical model \cite{riley2018}. The theoretical extension to the torque-given problem with $N=2$ flagella is presented in Appendix B where we highlight the similarities and differences between the two models. 

\subsection{Identical and axisymmetric flagellar propulsion}

 From  (\ref{eq:m04}), (\ref{eq:10})  and (\ref{eq:20}), we obtain the governing equations in a matrix form,

\begin{equation}
\mathcal{A}
\begin{pmatrix}
\bm{U} \\
\bm{\Omega} \\
\bm{\omega}^{(i)} \\
\end{pmatrix}
=
\begin{pmatrix}
\bm{0} \\
\bm{0} \\
-\bm{M}^{(i)}_{elast}
\end{pmatrix}
\label{eq:30a},
\end{equation}
where $A$ is a square matrix of order $2N+6$,  given by
\begin{equation}
\mathcal{A}=
\begin{pmatrix}
\displaystyle
C_D\mathsf{1}+\sum_{i=1}^{N}\int_0^{L^{(i)}}\mathsf{C}_i\,ds &
\displaystyle
\sum_{i=1}^{N}\int_0^{L^{(i)}}\mathsf{C}_i\cdot\mathsf{A}_i\,ds &
\displaystyle
\int_0^{L^{(i)}}\mathsf{C}_i\cdot\tilde{\mathsf{A}}_i\,ds \\
\displaystyle
\sum_{i=1}^{N}\int_0^{L^{(i)}}\mathsf{A}_i^T\cdot\mathsf{C}_i\,ds &
\displaystyle
C_R\mathsf{1}+\sum_{i=1}^{N}\int_0^{L^{(i)}}\mathsf{A}_i^T\cdot\mathsf{C}_i\cdot\mathsf{A}_i\,ds &
\displaystyle
\int_0^{L^{(i)}}\mathsf{A}_i^T\cdot\mathsf{C}_i\cdot\tilde{\mathsf{A}}_i\,ds \\
\displaystyle
\mathsf{P}_i\cdot\int_0^{L^{(i)}}\tilde{\mathsf{A}}_i^T\cdot\mathsf{C}_i\,ds &
\displaystyle
\mathsf{P}_i\cdot\int_0^{L^{(i)}}\tilde{\mathsf{A}}_i^T\cdot\mathsf{C}_i\cdot\mathsf{A}_i\,ds &
\displaystyle
\mathsf{P}_i\cdot\int_0^{L^{(i)}}\tilde{\mathsf{A}}_i^T\cdot\mathsf{C}_i\cdot\tilde{\mathsf{A}}_i\,ds 
\end{pmatrix}.
\end{equation}

We next decompose the flagellar rotation velocity vector into the components due to  the flagellar rotation and that due to the flagellar bending,  $\bm{\omega}^{(i)}=\bm{\omega}^{(i)}_t+\bm{\omega}^{(i)}_n$, with  $\bm{\omega}^{(i)}_t=\left(\bm{\omega}^{(i)}\cdot\bm{e}^{(i)}\right)\bm{e}^{(i)}$. In the rotation-given problem, 
$\omega^{(i)}_t=\bm{\omega}^{(i)}\cdot\bm{e}^{(i)}$ is a fixed value. 
For more convenience, we introduce $3\times3$ matrices to simplify the linear equations (\ref{eq:30a}) as
\begin{equation}
\begin{pmatrix}
\displaystyle
\mathsf{K}_{TT} & \mathsf{K}_{TR} & \mathsf{K}_{TF}^{(i)} & \\
\mathsf{K}_{RT} & \mathsf{K}_{RR} & \mathsf{K}_{RF}^{(i)} & \\
\mathsf{P}_i\cdot\mathsf{K}_{FT}^{(i)} & \mathsf{P}_i\cdot\mathsf{K}_{FR}^{(i)}  &  \mathsf{P}_i\cdot\mathsf{K}_{FF}^{(i)} 
\end{pmatrix}
\begin{pmatrix}
\bm{U} \\
\bm{\Omega} \\
\bm{\omega}^{(i)}_n 
\end{pmatrix}
=
\begin{pmatrix}
-\sum_{i=1}^{N}\mathsf{K}_{TF}^{(i)}\cdot\bm{\omega}^{(i)}_t \\
-\sum_{i=1}^{N}\mathsf{K}_{RF}^{(i)}\cdot\bm{\omega}^{(i)}_t \\
-\bm{M}^{(i)}_{elast}
-\mathsf{P}_i\cdot\mathsf{K}_{FF}^{(i)}\cdot\bm{\omega}^{(i)}_t
\end{pmatrix}.
\label{eq:31c}
\end{equation}

In order to proceed,  we assume that all  $N$ flagella are  identical and that they generate   axisymmetric propulsion around their long axes. This assumption is obviously satisfied for axisymmetric flagellar shapes   such as rods, but is also valid for  asymmetric shapes, including helices, provided that the time-scale of  flagellar rotation is much smaller than that of  elastic bending (i.e., $|\bm{\omega}_t|\gg |\bm{\omega}_n|$).  In practice, this assumption of axisymmetric propulsion  is  satisfied for swimming bacteria  \cite{lauga2006}. Indeed, the typical flagellar rotation frequency of flagellar filaments for  \textit{E.~coli}   is $\approx 100$~Hz, which is  faster than the typical frequency of cell rotation, $\approx 10$~Hz. We can therefore approximate the propulsion by a rotating helical flagellar filament by its time-averaged contribution, which is axisymmetric along the helix axis.

We introduce a reference frame, flagellum-fixed frames  and the rotation matrix, mapping  the laboratory reference frame, $\{ \bm{e}_x, \bm{e}_y,\bm{e}_z\}$,  onto the flagellum-fixed coordinates,  $\{ \bm{e}^{(i)}_x, \bm{e}^{(i)}_y,\bm{e}^{(i)}_z\}$  and denoted by $\mathsf{R}_i$, as schematically shown in Fig.~\ref{fig_model}. The axisymmetric conditions are satisfied if the quantity is unchanged under rotation around $\bm{e}^{(i)}_z$, where we choose the flagellum-fixed coordinates such that the $z$ axis ($\bm{e}^{(i)}_z$) coincides with $\bm{e}^{(i)}$ (See also Fig \ref{fig_model}).

For any second-rank  tensor written in the reference-frame as $\mathsf{K}$, its expression in the  body-fixed frame, $\mathsf{K}_i$, is given by
\begin{equation}
\mathsf{K}=\mathsf{R}_i\cdot\mathsf{K}_i\cdot\mathsf{R}^{-1}_i
\label{eq:50a}.
\end{equation}
We next assume that the tensor is axisymmetric around the vector $\bm{e}^{(i)}=\bm{e}^{(i)}_z$. This can be expressed by requiring invariance under rotation by any angle $\alpha\in[0,2\pi)$ around the axis, i.e.
\begin{equation}
\mathsf{K}=\mathsf{R}\left(-\alpha;\bm{e}^{(i)}\right)\cdot\mathsf{K}\cdot\mathsf{R}\left(\alpha;\bm{e}^{(i)}\right)
\label{eq:51},
\end{equation}
where
$\mathsf{R}\left(\alpha;\bm{e}^{(i)}\right)$ denotes the rotation matrix of angle $\alpha$ around $\bm{e}^{(i)}$. If the tensor $\mathsf{K}_i$ is represented as a matrix,  the axisymmetric property leads to  the matrix form, 
\begin{equation}
\mathsf{K}_i=\begin{pmatrix}
K_{11} & K_{12} & 0 \\
K_{21} & K_{22} & 0 \\
0 & 0 & K_{33} \\
\end{pmatrix}
\label{eq:52a},
\end{equation}
and the four upper-left entries can be decomposed into the symmetric and skew-symmetric parts,
\begin{equation}
\begin{pmatrix}
K_{11} & K_{12}  \\
K_{21} & K_{22}  \\
\end{pmatrix}
=\begin{pmatrix}
K_s & 0  \\
0 & K_s  \\
\end{pmatrix}
+\begin{pmatrix}
0 & K_{ss}  \\
-K_{ss} & 0  \\
\end{pmatrix}
\label{eq:52b},
\end{equation}
where $K_s$ and $K_{ss}$ are constants.
From the matrix form in (\ref{eq:52a}), note that we readily obtain the commutation relation, 
\begin{equation}
\mathsf{P}_i\cdot\mathsf{K}=\mathsf{K}\cdot\mathsf{P}_i
\label{eq:52c}.
\end{equation}

 The tensors $\tilde{\mathsf{A}}_i$ and $\mathsf{C}_i$    depend only on the flagellar shape  and if the flagellum shape is axisymmetric, these tensors are also axisymmetric. As a result  the tensor $\mathsf{K}^{(i)}_{FF}=\int\tilde{\mathsf{A}}^T_i\cdot\mathsf{C}_i\cdot\tilde{\mathsf{A}}_i\,ds$ is found to be axisymmetric, since $\mathsf{K}^{(i)}_{FF}
=\int\mathsf{R}\left(-\alpha;\bm{e}^{(i)}\right)\cdot\tilde{\mathsf{A}}^T_i\cdot\mathsf{C}_i\cdot\tilde{\mathsf{A}}_i\cdot\mathsf{R}\left(\alpha;\bm{e}^{(i)}\right)\,ds
=\mathsf{R}\left(-\alpha;\bm{e}^{(i)}\right)\cdot\mathsf{K}_{FF}^{(i)}\cdot\mathsf{R}\left(\alpha;\bm{e}^{(i)}\right)$.  If instead  the flagellar filaments are not rigorously axisymmetric, notably if they are  helical, then as long as the rotation velocity around their long axis is sufficiently large compared with the velocity scale during bending, we can approximately replace the second-rank tensors by their time-averages, and the same arguments thus follow.

Using the relation (\ref{eq:52c}),   equation (\ref{eq:31c}) can be simplified to
\begin{equation}
\begin{pmatrix}
\displaystyle
\mathsf{K}_{TT} & \mathsf{K}_{TR} & \mathsf{K}_{TF}^{(i)}  \\
\mathsf{K}_{RT} & \mathsf{K}_{RR} & \mathsf{K}_{RF}^{(i)} \\
\mathsf{P}_i\cdot\mathsf{K}_{FT}^{(i)} & \mathsf{P}_i\cdot\mathsf{K}_{FR}^{(i)}  &  \mathsf{K}_{FF}^{(i)} 
\end{pmatrix}
\begin{pmatrix}
\bm{U} \\
\bm{\Omega} \\
\bm{\omega}^{(i)}_n
\end{pmatrix}
=
\begin{pmatrix}
-\sum_{i=1}^{N}\mathsf{K}_{TF}^{(i)}\cdot\bm{\omega}^{(i)}_t \\
-\sum_{i=1}^{N}\mathsf{K}_{RF}^{(i)}\cdot\bm{\omega}^{(i)}_t \\
-\bm{M}^{(i)}_{elast}\\
\end{pmatrix}.
\label{eq:42a}
\end{equation}

Similar arguments for  $\mathsf{K}^{(i)}_{FF}$ enable us to show that the symmetric tensors $\mathsf{K}^{(i)}_{TF}=\mathsf{K}^{(i) T}_{FT}$  are also axisymmetric around the axis $\bm{e}^{(i)}$. However,  the tensors $\mathsf{K}^{(i)}_{RF}=\mathsf{K}^{(i) T}_{FR}$ are not necessarily axisymmetric, since $\mathsf{A}_i$ are not axisymmetric except when each $\bm{e}^{(i)}$ is parallel to  $\bm{x}^{(i)}$. We then decompose $\mathsf{A}_i=\mathsf{A}'_i+\tilde{\mathsf{A}}_i$. Noting that $\mathsf{A}'_i$ is independent of the flagellar shape, we rewrite as $\mathsf{K}_{RF}^{(i)}=\mathsf{A}_i^{'T}\cdot\mathsf{K}_{TF}^{(i)}+\mathsf{K}_{FF}^{(i)}$ so that the tensors $\tilde{\mathsf{A}}_i$, $\mathsf{K}^{(i)}_{TF}$ and $\mathsf{K}^{(i)}_{FF}$ are axisymmetric around the vector $\bm{e}^{(i)}$.  For convenience, we write $\mathsf{A}_i^{'T}\cdot\mathsf{K}_{TF}^{(i)}=\mathsf{K}_{RF}^{'(i)}$.

We next employ the  assumption that all flagellar filaments have identical  propulsion, which guarantees that the axisymmetric tensor, $\mathsf{K}$, can be expressed as  $\mathsf{K}=\mathsf{R}_i\cdot\mathsf{K}^{(0)}\cdot\mathsf{R}_i^{-1}$, where the tensor $\mathsf{K}^{(0)}$ is common to all filaments. We also write the rotation of the flagellar angular velocity in the flagellum-fixed frame as $\bm{\omega}^{(i)}_t=\mathsf{R}_i\cdot\tilde{\bm{\omega}}^{(i)}_t$; here  $\tilde{\bm{\omega}}^{(i)}_t$ is parallel to $\bm{e}_z$ and we can thus write $\tilde{\bm{\omega}}^{(i)}_t=\tilde{\omega}^{(i)}_t\bm{e}_z$, where the value of  $\tilde{\omega}^{(i)}_t$ is prescribed. Similarly, the elastic torque in the flagellum-fixed frame is written as $\bm{M}^{(i)}_{elast}=\mathsf{R}_i\cdot\tilde{\bm{M}}^{(i)}_{elast}$. Since both $\mathsf{K}^{(i)}_{TF}$ and $\mathsf{K}^{(i)}_{FF}$ are axisymmetric, we can rewrite the linear equations (\ref{eq:42a}) as
\begin{equation}
\begin{pmatrix}
\displaystyle
\mathsf{K}_{TT} & \mathsf{K}_{TR} & \mathsf{R}_i\cdot\mathsf{K}_{TF}^{(0)} \\
\mathsf{K}_{RT} & \mathsf{K}_{RR} & \mathsf{K}_{RF}^{(i)}\cdot\mathsf{R}_i\\
\mathsf{R}_i\cdot\mathsf{P}_0\cdot\mathsf{K}_{FT}^{(0)}\cdot\mathsf{R}_i^{-1} & \mathsf{P}_i\cdot\mathsf{K}_{FR}^{(i)}  &  \mathsf{R}_i\cdot\mathsf{K}_{FF}^{(0)} 
\end{pmatrix}
\begin{pmatrix}
\bm{U} \\
\bm{\Omega} \\
\tilde{\bm{\omega}}^{(i)}_n
\end{pmatrix}
=
\begin{pmatrix}
-\sum_{i=1}^{N}\mathsf{R}_i\cdot\mathsf{K}_{TF}^{(0)}\cdot\tilde{\bm{\omega}}^{(i)}_t \\
-\sum_{i=1}^{N}\left(\mathsf{K}_{RF}^{'(i)}\cdot\mathsf{R}_i+\mathsf{R}_i\cdot\mathsf{K}_{FF}^{(0)}\right)\cdot\tilde{\bm{\omega}}^{(i)}_t \\
-\mathsf{R}_i\cdot\tilde{\bm{M}}^{(i)}_{elast}
\end{pmatrix},
\label{eq:42b}
\end{equation}
where $\mathsf{P}_0$ is the projection onto the $x-y$ plane, defined as $\mathsf{P}_i=\mathsf{R}_i\cdot\mathsf{P}_0\cdot\mathsf{R}_i^{-1}$. With additional rotational matrix in the bottom $N$ rows, this can be simplified to
\begin{align}
\begin{pmatrix}
\displaystyle
\mathsf{K}_{TT} & \mathsf{K}_{TR} & \mathsf{R}_i\cdot\mathsf{K}_{TF}^{(0)} \\
\mathsf{K}_{RT} & \mathsf{K}_{RR} & \mathsf{K}_{RF}^{(i)}\cdot\mathsf{R}_i \\
\mathsf{P}_0\cdot\mathsf{K}_{FT}^{(0)}\cdot\mathsf{R}_{i}^{-1}  & \mathsf{P}_0\cdot\mathsf{R}_i^{-1}\cdot\mathsf{K}_{FR}^{(i)} & \mathsf{K}_{FF}^{(0)} 
\end{pmatrix}
\begin{pmatrix}
\bm{U} \\
\bm{\Omega} \\
\tilde{\bm{\omega}}^{(i)}_n 
\end{pmatrix} 
=
\begin{pmatrix}
-\bm{F}_{prop} \\
-\bm{M}_{prop} \\
-\tilde{\bm{M}}^{(i)}_{elast}
\end{pmatrix}.
\label{eq:42c}
\end{align}

Here, we have used the fact that $\mathsf{P}_0$ commutes with  $\mathsf{K}^{(0)}_{FT}$ and $\tilde{\mathsf{K}}^{(0)}_{FR}$  and that $\mathsf{P}_0\cdot\tilde{\bm{\omega}}_n^{(i)}=\tilde{\bm{\omega}}_n^{(i)}$.
The first and second entries of the right-hand side of (\ref{eq:42c}) give the effective force and torque for the entire body, corresponding to propulsive force and torques, 
\begin{eqnarray}
\bm{F}_{prop}&=&\left(\sum_{i=1}^{N}\tilde{\omega}^{(i)}_t\mathsf{R}_i\right)\cdot\mathsf{K}_{TF}^{(0)}\cdot\bm{e}_z,
\label{eq:43a} \\
\bm{M}_{prop}&=&\sum_{i=1}^{N}\mathsf{K}_{RF}^{'(i)}\cdot\bm{\omega}^{(i)}_t+\left(\sum_{i=1}^N\tilde{\omega}^{(i)}_t\mathsf{R}_i\right)\cdot\mathsf{K}_{FF}^{(0)}\cdot\bm{e}_z
\label{eq:43b} .
\end{eqnarray}

For a symmetric configuration such that $\left(\sum_{i=1}^N \tilde{\omega}^{(i)}_t\mathsf{R}_i\right)\cdot\bm{e}_z=\bm{0}$, only the first term of $\bm{M}_{prop}$ contributes to the  motion of the cell.  If we further assume that the reference flagellar orientation is  normal to the sphere surface, i.e. $\bm{e}^{(i)}_z(t=0)=\bm{n}^{(i)}$, we readily obtain $\tilde{\bm{M}}^{(i)}_{elast}=\bm{0}$  and $\bm{M}_{prop}=\bm{0}$.
Therefore, we have stationary solutions satisfying $\bm{U}=\bm{\Omega}=\tilde{\bm{\omega}}_n^{(i)}=\bm{0}$ for the symmetric-filament configurations.

From the  axisymmetric propulsion of the flagellar filaments  and  using (\ref{eq:52a})-(\ref{eq:52b}), we may obtain the general form of the tensors
\begin{equation}
\mathsf{K}_C^{(0)}=
\begin{pmatrix}
k_C & 0   & 0    \\
0   & k_C & 0    \\
0   & 0   & K_C  \\
\end{pmatrix}
,~
\mathsf{K}_{TF}^{(0)}=
\begin{pmatrix}
 k_D   & k_T  & 0    \\
-k_T&   k_D    & 0     \\
0   & 0    & K_T  \\
\end{pmatrix}
,~
\mathsf{K}_{FF}^{(0)}=
\begin{pmatrix}
k_F & 0   & 0    \\
0   & k_F & 0    \\
0   & 0   & K_F  \\
\end{pmatrix},
\label{eq:90a1}
\end{equation}
where the constants, $k_C, K_C, k_T, k_F$, are all negative due to the negative definiteness of the Stokes resistance tensors  and  $K_T$ and $K_F$ are taken to be  negative so that a positive value of $\tilde{\omega}_t$ generates   flagellar force  towards the cell body   (as is the case for the majority of flagellated bacteria).  From the definitions of $\mathsf{K}^{(0)}_C$ and $\mathsf{K}^{(0)}_{FF}$, these matrices are symmetric, thus the expressions in (\ref{eq:90a1}) follow.  The expression for $\mathsf{K}_{TF}^{(0)}$ reflects the skew-symmetric property of $\tilde{A}_i$, and thus the diagonal component, $k_D$, is typically very small for  swimming bacteria. The diagonal component of the matrix can be either positive or negative depending on the  chirality of flagellar filaments. Using the Appendix in Ref.~\cite{lauga2006}, $k_D$ can be estimated for a helical flagellum as $k_D\sim C_N bL\epsilon$ with $b$ is the diameter of the helix and $\epsilon=2\pi b/\lambda$, where $\lambda $ is the pitch of the helix. The ratio of diagonal to off-diagonal component is thus given by $k_D/k_T\sim \epsilon (b/L)\sim 10^{-2}$, always a small number for a bacterium such as \textit{E.~coli}  \cite{lauga2006}. Note that $k_D$ is identically zero for rod-like flagella. 

 We now proceed with the linear stability problem  neglecting the chirality effects (i.e.~setting $k_D=0$)  and will then incorporate chirality back in  Section~\ref{Sec:chilarity}.
\subsection{Linear stability}

\begin{figure}[t]
\centering\includegraphics[width=13cm]{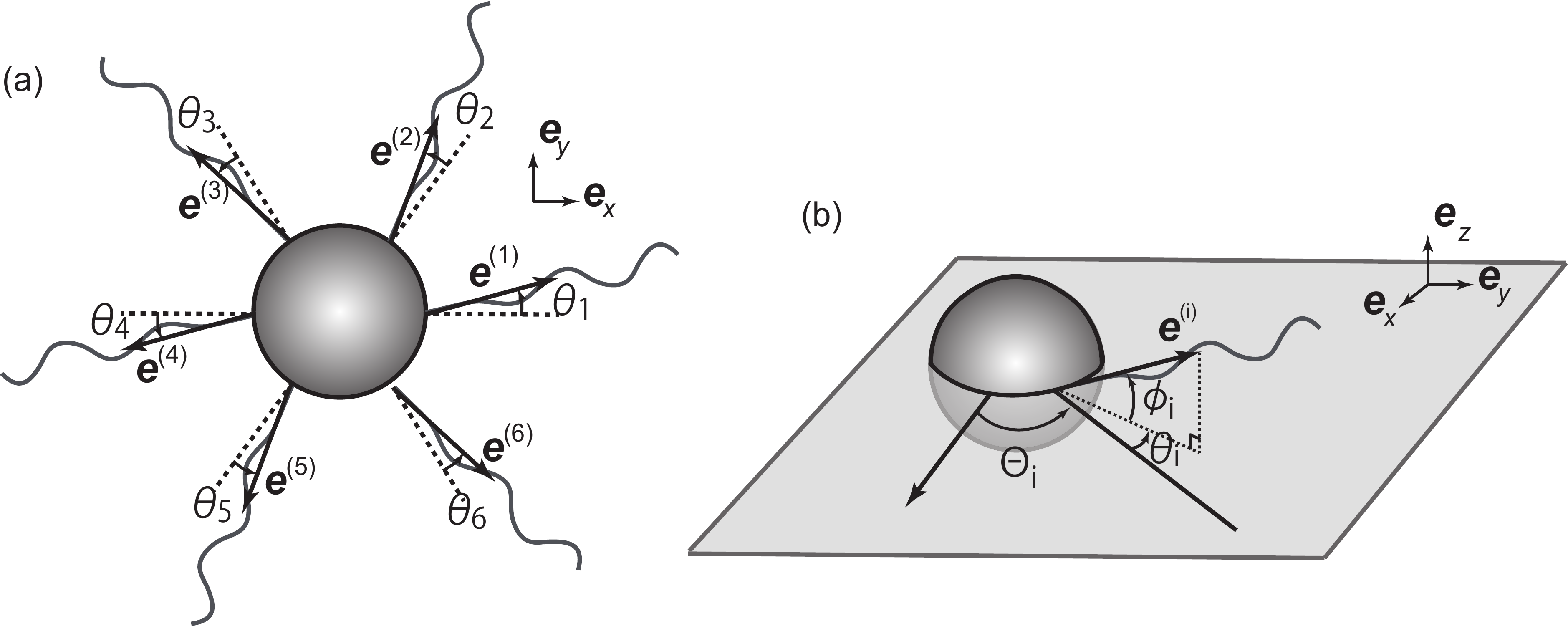}
\caption{(a) Schematic representation of a model bacterium with $N$ identical flagella symmetrically attached in the $x-y$ plane. (b) Angles specifying the orientation of a flagellum. The equilibrium orientation of $i$th flagellum is given by the rotation of angle $\Theta_i$ from the $+x$ axis. Small deformations from the equilibrium are measured by the two angles, $\theta_i$ and $\phi_i$, which correspond to the in-plane and out-of-plane angles, respectively.}
\label{fig_Nflag}
\end{figure}

{In this section, we  assume the in-plane flagellar configuration is that of a regular polygon,  and then formulate the linear stability problem around the equilibrium.} The example case of $N=6$ is schematically shown in Fig.~\ref{fig_Nflag}(a) with notation.  As shown above, with a symmetric flagellar configuration we obtain stationary solutions for the motion of a bacterium with $N (\geq 2)$ identical axisymmetric flagella. We assume that the  angular rotations of each flagellum, $\bm{\omega}^{(i)}_t$, are identical  and that each flagellar filament is connected perpendicularly to the cell body surface and located in one plane that we define as $x-y$ plane (see Fig.~\ref{fig_Nflag}a). When the flagellar configuration forms an $N$-sided regular polygon, the bacterial motion is found to be stationary.

We now consider the linear stability around this equilibrium, where the orientation of $i$th flagellum is given by rotation of angle $\Theta_i$ from the $+x$ axis in the $x-y$ plane  where $\Theta_i=2\pi(i-1)/N$ (see Fig.~\ref{fig_Nflag}b). The disturbance from the equilibrium is denoted by two angles for each flagellum. We introduce the in-plane and out-of-plane angle displacements for the $i$th flagellum as $\theta_i$ and $\phi_i$, as shown in Fig.~\ref{fig_Nflag}(a, b).  In the linear stability regime, both angles are assumed to be sufficiently small, i.e.~$|\theta_i|, |\phi_i| \ll 1$. 
We   write the flagellar bending rotation as $\tilde{\bm{\omega}}^{(i)}_n=(d\theta_i/dt)\bm{e}^{(i)}_x+(d\phi_i/dt)\bm{e}^{(i)}_y$,   assume a constant flagellar rotation around the axisymmetric axis, $\tilde{\bm{\omega}}^{(i)}_t=\omega_0$, and also assume that the torque spring constants are identical  for all  flagella, i.e.~$\kappa^{(i)}=\kappa$.

We then proceed to obtain the expressions for the cell dynamics around the equilibrium configuration. The rotation matrix, $\mathsf{R}_i$ is obtained by  combining the two rotations,
\begin{eqnarray}
\mathsf{R}_i&=&\mathsf{R}(\phi_i;\bm{e}_y)\cdot\mathsf{R}(\Theta_i+\theta_i; \bm{e}_x)\cdot\begin{pmatrix}
0 & 0 & 1 \\
0 & -1&0 \\
1 & 0 & 0
\end{pmatrix} \nonumber \\
&\simeq& 
\begin{pmatrix}
0 & \sin\Theta_i & \cos\Theta_i \\
0 & -\cos\Theta_i & \sin\Theta_i \\
1 & 0& 0
\end{pmatrix}
+\theta_i
\begin{pmatrix}
0 & \cos\Theta_i &  -\sin\Theta_i \\
0 & \sin\Theta_i & \cos\Theta_i \\
0 & 0& 0
\end{pmatrix}
+\phi_i\begin{pmatrix} 
-\cos\Theta_i & 0 & 0 \\
-\sin\Theta_i & 0  &0\\
0 & 0 & 1
\end{pmatrix}
\label{eq:90a0},\quad\quad
\end{eqnarray}
where $\mathsf{R}(\phi_i;\bm{e}_y)$ denotes the rotation matrix with angle $\phi_i$ around the orientation $\bm{e}_y$. 
Using the expressions (\ref{eq:90a1}), we directly obtain the effective force, 
\begin{equation}
\bm{F}_{prop}=
\left(\sum_{i=1}^N \omega_0\mathsf{R}_i\right)\cdot\mathsf{K}^{(0)}_{TF}\cdot\bm{e}_z= \omega_0 K_T\left( - \sum_{i=1}^N \theta_i\sin\Theta_i\bm{e}_x+ \sum_{i=1}^N\theta_i\cos\Theta_i\bm{e}_y+\sum_{i=1}^N \phi_i\bm{e}_z\right)
\label{eq:90a},
\end{equation}
and similarly the effective torque,
\begin{align}
\bm{M}_{prop}=&
\omega_0K_T\left( \sum_{i=1}^N \phi_i\sin\Theta_i\bm{e}_x -\sum_{i=1}^N \phi_i \cos\Theta_i\bm{e}_y+\sum_{i=1}^N \theta_i \bm{e}_z \right) \nonumber \\
&+
\omega_0K_F\left( - \sum_{i=1}^N \theta_i\sin\Theta_i\bm{e}_x+ \sum_{i=1}^N\theta_i\cos\Theta_i\bm{e}_y+\sum_{i=1}^N \phi_i\bm{e}_z\right)
\label{eq:90b1}.
\end{align}

We then proceed to compute the force and torque generated by the in-plane flagellar bending, $\tilde{\bm{\omega}}_n^{(i)}=(d\theta_i/dt)\bm{e}_x$. The force generated by each flagellar filament is $\mathsf{R}_i\cdot\mathsf{K}_{TF}^{(0)}\cdot\tilde{\bm{\omega}}_n^{(i)}\simeq k_T(d\theta_i/dt) (-\sin\Theta_i\bm{e}_x+\cos\Theta_i\bm{e}_y)$. Similarly, the torque generated can also be computed, using the decomposition 
$\mathsf{K}_{RF}^{(i)}\cdot\mathsf{R}_i\cdot\tilde{\bm{\omega}}_n^{(i)}=
\left( \mathsf{A}_i^{'T}\cdot\mathsf{R}_i\cdot\mathsf{K}_{TF}^{(0)}+\mathsf{R}_i\cdot\mathsf{K}_{FF}^{(0)}\right)\cdot\tilde{\bm{\omega}}_n^{(i)}
=(k_T+k_F)(d\theta_i/dt)\bm{e}_z
$. 

The force generated  by the out-of-plane flagellar bending, $\tilde{\omega}_n^{(i)}=(d\phi_i/dt)\bm{e}_y$, is also computed as $\mathsf{R}_i\cdot\mathsf{K}_{TF}^{(0)}\cdot (d\phi_i/dt)\bm{e}_y=k_T(d\phi_i/dt)\mathsf{R}_i\cdot\bm{e}_x\simeq k_T(d\phi_i/dt)\bm{e}_z$ . Similarly, we obtain the torque generated by the bending, $(d\phi_i/dt)\bm{e}_y$, as $\mathsf{K}_{RF}^{(i)}\cdot\mathsf{R}_i\cdot(d\phi_i/dt)\bm{e}_y=(d\phi_i/dt)(\mathsf{A}_i^{'T}\cdot\mathsf{R}_i\cdot\mathsf{K}_{TF}^{(0)}+\mathsf{R}_i\cdot\mathsf{K}_{FF}^{(0)})\cdot\bm{e}_y=(d\phi_i/dt)(k_T\mathsf{A}_i^{'T}\cdot\mathsf{R}_i\cdot\bm{e}_x+k_F\mathsf{R}_i\cdot\bm{e}_y)=(d\phi_i/dt)(k_T+k_F)(\sin\Theta_i\bm{e}_x-\cos\Theta_i\bm{e}_y)$.

A lengthy but straightforward calculation then  leads to the matrix expression for the motion around the equilibrium, captured by a $2N+6$ dimensional linear problem,
\begin{equation}
\mathcal{A}\bm{\Phi}=\bm{b}
\label{eq:90c},
\end{equation}
where 
$\bm{\Phi}=(U_x, U_y, U_z, \Omega_x, \Omega_y, \Omega_z,  \dot{\theta}_1,\cdots,\dot{\theta}_N,\dot{\phi}_1,\cdots,\dot{\phi}_N)^T$  and 
\begin{equation}
\bm{b}=(-\bm{F}_{prop}^T,-\bm{M}_{prop}^T,\kappa\theta_1,\cdots,\kappa\theta_N,\kappa\phi_1,\cdots,\kappa\phi_N)^T
\label{eq:90c2}.
\end{equation}

The detailed derivations leading to the expressions for the square matrix $\mathcal{A}$ can be found in  Appendix A  and the results are
\begin{equation}
\mathcal{A}=
\begin{pmatrix}
C_{D1} &  0 & 0 &
0 & 0 & 0 &
\bm{C}_{TF1}^T & \bm{0} \\
0 & C_{D2} & 0 &
0 & 0 & 0 &
\bm{C}_{TF2}^T & \bm{0} \\
0 & 0 & C_{D3} &
0 & 0 & 0 &
\bm{0} & \bm{C}_{TF3}^T \\
0 & 0 & 0 &
C_{R1} &  0 & 0 &
\bm{0} & \bm{C}_{RF1}^T \\
0 & 0 & 0 &
0 & C_{R2} & 0 &
\bm{0} & \bm{C}_{RF2}^T  \\
0 & 0 & 0 &
0 & 0 & C_{R3} &
\bm{C}_{RF3}^T & \bm{0} \\
\bm{C}_{TF1} & \bm{C}_{TF2} & \bm{0} &
\bm{0} & \bm{0} & \bm{C}_{RF3} &
k_F\bm{1}_N & \bm{0}_N \\
\bm{0} & \bm{0} & \bm{C}_{TF3} &
\bm{C}_{RF1} & \bm{C}_{RF2} & \bm{0} &
\bm{0}_N& k_F\bm{1}_N \\
\end{pmatrix}
\label{eq:75},
\end{equation}
where the constants in the matrix are  
 \begin{align}
 C_{D1}=&C_D+\sum_{i=1}^N \left( K_C\cos^2\Theta_i+k_C\sin^2\Theta_i\right)
 \label{eq:76a}, \\
  C_{D2}=&C_D+\sum_{i=1}^N \left( K_C\sin^2\Theta_i+k_C\cos^2\Theta_i\right)
 \label{eq:76a2}, \\
 C_{D3}=&C_D+Nk_C
  \label{eq:76b}, \\
 C_{R1}=&C_R+\sum_{i=1}^N \left( (k_C+2k_T+k_F)\sin^2\Theta_i+K_F\cos^2\Theta_i\right)
  \label{eq:76c}, \\
   C_{R2}=&C_R+\sum_{i=1}^N \left( (k_C+2k_T+k_F)\cos^2\Theta_i+K_F\sin^2\Theta_i\right)
  \label{eq:76c2},\\
  C_{R3}=&C_R+N(k_C+2k_T+k_F)
  \label{eq:76d},
 \end{align} 
 and the $N$ dimensional vectors are given by
 \begin{align}
\bm{C}_{TF1}=-k_T(\sin\Theta_1,\cdots,-\sin\Theta_N)^T,
&~ \bm{C}_{RF1}=(k_T+k_F)(\sin\Theta_1,\cdots,\sin\Theta_N)^T,  \\ 
\bm{C}_{TF2}=k_T(\cos\Theta_1,\cdots,\cos\Theta_N)^T,
&~\bm{C}_{RF2}=-(k_T+k_F)(\cos\Theta_1,\cdots,\cos\Theta_N)^T,
 \\
\bm{C}_{TF3}=(k_T,\cdots,k_T)^T,
&~\bm{C}_{RF3}=(k_T+k_F,\cdots,k_T+k_F)^T,
\end{align}
and we used   $\bm{1}_N$ and $\bm{0}_N$ to denote the identity and zero matrices of order $N$.
 In the following sections, we will consider the stability of this linear system.

\section{Instability of bacteria with $N=2$ flagella}
\label{sec:N=2}

\begin{figure}[t]
\centering\includegraphics[width=13cm]{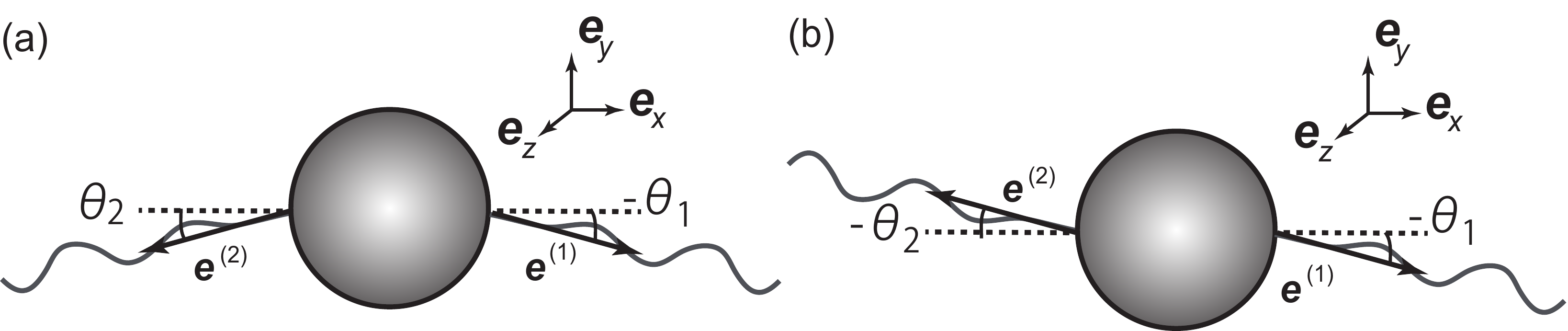}
\caption{Schematic pictures of the two modes of flagellar configuration for the $N=2$ case. (a) Translation mode when the  two flagella are in a mirror-symmetric  configuration. (b) Rotation mode when the two  flagella are in a  point-symmetric configuration.}
\label{fig_twoflag}
\end{figure}

We start with analysing  the $N=2$ case. Substituting $N=2$ in the expressions (\ref{eq:90c})- (\ref{eq:75}), we   obtain a 10-dimensional linear system (Fig.~\ref{fig_twoflag}).
With the geometric symmetry of the system, we readily find that the $x$ component of the effective force and torque vanish  and thus $U_x=\Omega_x=0$ follow. The system is then reduced to an 8-dimensional linear system.
 Introducing the variables   $\theta_+=\theta_1+\theta_2$, $\theta_-=\theta_1-\theta_2$, $\phi_+=\phi_1+\phi_2$  and $\phi_-=\phi_1-\phi_2$, we can rewrite the linear system (\ref{eq:90c}) into 4 blocks of 2$\times$2 matrices as
\begin{align}
\begin{pmatrix}
C_D' & k_T \\
2k_T & k_F
\end{pmatrix}
\begin{pmatrix}
U_y \\
\dot{\theta}_-
\end{pmatrix}
&=
\begin{pmatrix}
-\omega_0K_T \theta_- \\
\kappa\theta_-
\end{pmatrix},
\label{eq:81d01} \\
\begin{pmatrix}
 C_D' & k_T    \\
 2k_T & k_F
\end{pmatrix}
\begin{pmatrix}
U_z \\
\dot{\phi}_+
\end{pmatrix}
&=
\begin{pmatrix}
-\omega_0K_T \phi_+ \\
\kappa\phi_+
\end{pmatrix}\label{eq:81d02}, \\
\begin{pmatrix}
C_R'  &  -(k_T+k_F) \\
-2(k_T+k_F) & k_F
\end{pmatrix}
\begin{pmatrix}
\Omega_y\\
\dot{\phi}_-\\
\end{pmatrix}
&=
\begin{pmatrix}
\omega_0K_T\phi_-  - \omega_0K_F\theta_-\\
\kappa\phi_-\\
\end{pmatrix},
\label{eq:81d03} \\
\begin{pmatrix}
 C_R'     & k_T+k_F  \\
2(k_T+k_F)      & k_F
\end{pmatrix}
\begin{pmatrix}
\Omega_z\\
\dot{\theta}_+
\end{pmatrix}
&=
\begin{pmatrix}
-\omega_0K_T\theta_+ -\omega_0K_F\phi_+\\
\kappa\theta_+\\
\end{pmatrix},
\label{eq:81d04}
\end{align}
where we have defined $C_D'=C_D+2k_C$ and $C_R'=C_R+2k_C+4k_T+2k_F$  and where the dot symbol indicates the time derivative of the angle variable.

Solving each 2-dimensional problem yields  ordinary differential equations with respect to the four angle variables as
\begin{equation}
\frac{d}{dt}
\begin{pmatrix}
\theta_+\\
\phi_+\\
\phi_-\\
\theta_-
\end{pmatrix}
=\begin{pmatrix}
A_R & A_{RT}& 0 & 0\\
0 & A_T& 0 & 0\\
0 &0 & A_R& -A_{RT} \\
0&0&0& A_T
\end{pmatrix}\begin{pmatrix}
\theta_+\\
\phi_+\\
\phi_-\\
\theta_-
\end{pmatrix}
\label{eq:81d},
\end{equation}
where the expressions of $A_T, A_R, A_{RT}$ are given by 
\begin{align}
A_T&=\Delta_D^{-1}(2\omega_0|K_Tk_T| - \kappa|C_D'|), 
\label{eq:81d2a} \\
A_R&=\Delta_R^{-1}(2\omega_0|K_T(k_T+k_F)-\kappa|C_R'|), \label{eq:81d2b}\\
A_{RT}&=\Delta_R^{-1}(2\omega_0|K_F(k_T+k_F)|,
\label{eq:81d2c}
\end{align}
and where the two determinants,
$\Delta_D=C_D'k_F-2k_F^2$ and
$\Delta_R=C_R'k_F-2(k_T+k_F)^2$, are  positive as a result of the  negative-definiteness of the Stokes resistance  matrices.

We first consider the simple  case where $K_F=0$ where the flagellar filaments produce propulsion but  no torque. This assumption follows for a rod-like active filament as in 
Ref.~\cite{riley2018}. 
Under this assumption, the matrix (\ref{eq:81d}) is diagonal, since $A_{RT}=0$  and with eigenvalues $A_T$ and $A_R$. The angles $\theta_-$ and $\phi_+$ correspond to the translation modes in the $y$ and $z$ directions respectively (Fig.~\ref{fig_twoflag}a).
The critical flagellar angular velocity, $\omega_{0T}$, above which the translation mode becomes unstable is given by
\begin{equation}
\omega_{0T}=\frac{|C_D+2k_C|}{2|K_Tk_T|}\kappa
\label{eq:81e1},
\end{equation}
which is  positive. When $\omega_0>0$, the flagellar filaments exert propulsive forces pushing on the cell body, whereas they pull on the organism when  $\omega_0 <0$. We therefore obtain  that the translation instability only occurs for  flagellar filaments in the pushing mode and with  sufficiently large propulsive magnitude (or, for a fixed propulsion, with a sufficiently flexible hook). 
This can be compared to the arguments put forward in 
Ref.~\cite{riley2018}, where the force by a rod-like flagellum corresponds to the product $\omega_0K_T$. Note that here the translation can occur towards an arbitrary direction in the $y-z$ plane.

The eigenvectors corresponding  to the eigenvalue $A_R$ are linear combinations of the angles $\theta_+$ and $\phi_-$ and characterise a rotation mode around $z$ and $y$ axis, respectively (Fig.~\ref{fig_twoflag}b).  Although the rotation mode has not been examined in Ref.~\cite{riley2018} (that study assumed mirror-image symmetry in their theoretical description), the rotation instability can occur above a second critical flagellar angular velocity, $\omega_{0R}$,  given by 
\begin{equation}
\omega_{0R}=\frac{|C_R+2k_C+4k_T+2k_F|}{2|K_T(k_T+k_F)|}\kappa
\label{eq:81e2}.
\end{equation}
Note that this critical value is positive and therefore  the rotation instability also occurs only when the flagellar filaments in the pushing mode.

In  the case where $K_F\neq 0$,   each flagellar filament generates both torque and force. Again, we obtain the same eigenvalues of the matrix, $A_R, A_T$, as in the problem of the flagella without torque generation. Moreover, the eigenvectors associated with the  eigenvalue $A_R$ are the same as above (pure rotation move). As a difference, however, the eigenvectors for  the eigenvalue $A_T$ are  combinations of the four angles  and the induced motion is found to be a  translation in $y-z$ plane  combined with a rotation around the translation direction. This unstable mode   corresponds to the case of bacterial flagellar bundling via the elastohydrodynamic instability as observed experimentally and reproduced using   numerical simulations in Ref.~\cite{riley2018}.

\subsection{Most unstable mode}
\label{Sec:unst2}

\begin{figure}[t]
\centering\includegraphics[width=14cm]{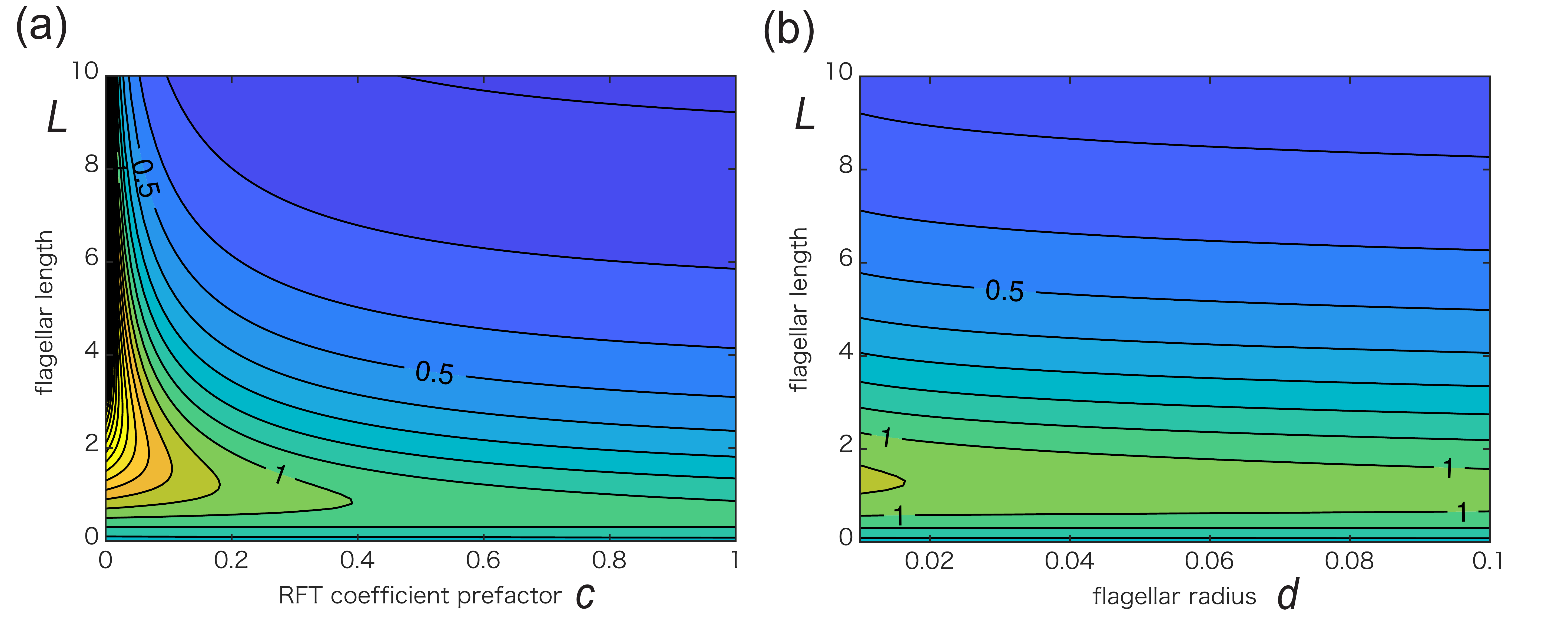}
\caption{Ratio between the two  values of angular velocities for the translation and rotation modes, $r=\omega_{0T}/\omega_{0R}$, as a function of  flagellar parameters. (a) Iso-values of $r$ as a function of the RFT coefficient prefactor $c$ and the flagellar length $L$. (b) Iso-values of $r$ as a function of flagellar radius $d$ and the flagellar length $L$.}
\label{fig_mostunst}
\end{figure}

When the angular velocity  of each flagellar filament  is positive, the two modes can become unstable. However, only the most unstable mode from the linear stability theory is likely to be observed in practice  and we now consider which one of the two modes becomes unstable first. 

Let us denote the length of the identical flagella by $L$. Using dimensional analysis allows to obtain  order-of-magnitude estimates for the dependence with $L$ of the  constants from the flagellar shape, namely  $k_C=O(L)$, $k_T=O(L^2)$ and $k_F=O(L^3)$. Noting that $C_D=-6\pi$ and $C_R=-8\pi$, we can then estimate the size of the ratio $r$ between the two critical angular velocities, $r\equiv\omega_{0T}/\omega_{0R}$, as $r\sim C_D/C_R=3/4$ when $L\ll 1$  and $r=O(L^{-1})$ when $L\gg 1$. In both limits, we see that the critical value for the translation instability is smaller  and thus it is the one which would be observed. 
 
 For further discussions in the intermediate region of $L$, we consider rod-like flagellar filaments of radius $d$ and length $L$. 
We introduce a positive constant $c$ such that  $C_n=-4\pi\mu c$ is the normal drag coefficient, $c$ is given by $c=(\log(2L/d)+0.5)^{-1}$ \cite{lauga2009}. Within  resistive force theory, the coefficients are given by $k_C=C_nL$, $k_T=(C_n/2)L^2$ and $k_F=(C_n/3)L^3$  and we can compute numerically the ratio $r$ for different flagellar parameters, $c$ and $L$, or $d$ and $L$, with results plotted in Fig.~\ref{fig_mostunst}.

In Fig.~\ref{fig_mostunst}(a), iso-values of the frequency  ratio $r$ is  first shown for different values $c$ and $L$  and we find that in the intermediate region   $L\sim1$ the rotation mode can be more unstable, although the translational mode is always more unstable above a critical value of $c\approx0.4$.  The same plot is then shown with the flagellar radius $d$ in the horizontal axis of Fig.~\ref{fig_mostunst}(b), indicating that the rotation instability would occur first in a robust range of flagellar radius if $L\sim1$. 

Typical sizes of the cell body and the flagellar filaments of \textit{E.~coli} are $\approx 1~\mu$m and $\approx 10~\mu$m \cite{lauga2009}, leading to a large  non-dimensional flagellar length, $L\approx10$,  and thus the translation mode is predicted to be the experimentally-observable one. Similarly, the typical size of the dimensionless flagellar radius and the RFT coefficient prefactor are given by $d\approx 0.02$ and $c\approx 0.13$, respectively,  and  a cell equipped with short flagellar filaments could therefore, in theory, undergo the rotational instability without net locomotion.

We next consider a helical flagellum whose shape is characterised by the helix angle $\Psi$ between the  local flagellar tangent vector, $\bm{t}^{(i)}$  and the axis of the helix, $\bm{e}^{(i)}_z$. The   tangent vector can be expressed by $\bm{t}^{(i)}=\cos\Psi\bm{e}^{(i)}_z+\sin\Psi(-\sin\beta\bm{e}^{(i)}_x+\cos\beta\bm{e}^{(i)}_y)$ with the angle $\beta$ in the range $\beta\in[0,2\pi)$. For the flagellar filament moved along the  $\bm{e}^{(i)}_x$ axis with velocity $u$, direct computations from (\ref{eq:m01}) gives  the local drag force,
\begin{equation}
d\bm{F}\cdot\bm{e}^{(i)}_x=C_n\left[ 1-(1-\gamma)\sin^2\Psi\sin^2\beta\right]u
\label{eq:82a},
\end{equation}
where $\gamma=C_t/C_n$ is the ratio of the tangential and normal coefficients from   resistive force theory. If the timescale of flagellar rotation is sufficiently faster than that of bending, we can approximate the local force by its time-averaged value, which is obtained  averaging over the angle parameter $\beta$. We thus obtain $d\bm{F}\cdot\bm{e}^{(i)}_x=C'_n u$, with the effective normal drag coefficient $C'_n =-4\pi\mu c'$ as
\begin{equation}
C'_n=\left[1-\frac{1}{2}(1-\gamma)\sin^2\Psi\right]C_n
\label{eq:82b},
\end{equation}
enabling us to follow the same arguments for the rod-like flagella by simply replacing $C_n$ by $C'_n$,  or $c$ by $c'$.
Typical values of $\gamma$ and $\Psi$ for \textit{E.~coli}  are $\gamma\approx 0.7$\cite{chattopadhyay2006} and  $\Psi\approx 30$ deg.\cite{spagnolie2011}, resulting in  the effective drag coefficient, and thus the effective value of $c$, to  be  similar  to the case of a rod ($c'\approx0.13$). As before, the translation mode is therefore more unstable for a bacterium with typical sizes, though there is a region where the rotation mode can be more unstable when $L\sim 1$.  Note however that this assumes that  the  helical structure of the filament is neglected, a modelling assumption which   is corrected in Section \ref{Sec:chilarity}.

\section{Instability of bacteria with $N\geq 3$ flagella}
\label{sec:generalN}

We now proceed to study the   linear stability problem in  the general  case of $N$ flagellar filaments. When $N\geq3$, the drag coefficients (\ref{eq:76a}) - (\ref{eq:76d}) can be simplified to
 \begin{align}
 C_{D1}=C_{D2}=C_D+(N/2)\left( K_C+k_C\right)
,&~
 C_{D3}=C_D+Nk_C
  \label{eq:96b}, \\
 C_{R1}=C_{R2}=C_R+(N/2) \left( k_C+2k_T+k_F+K_F\right)
,&~
  C_{R3}=C_R+N(k_C+2k_T+k_F)
  \label{eq:96d},
 \end{align}
 where we have used the equalities,
\begin{equation}
\sum_{i=1}^N \cos^2\left(\frac{2\pi(i-1)}{N}\right)=\sum_{i=1}^N \sin^2\left(\frac{2\pi(i-1)}{N}\right)=\frac{N}{2}
\label{eq:79a},
\end{equation}
 which are satisfied when $N\geq 3$.
 We first address the  $N=3$ and $N=4$ cases to allow for intuition on the mathematical structure of the solution, before proceeding to the general $N$ case.
 
 \subsection{Cell with $N=3$ flagella}
 
 When $N=3$, the number of the angular variables is 6, which is equal to the  number of  degrees of freedom for the rigid motion of the whole cell. As in the previous section, we   partially diagonalise the matrix (\ref{eq:75})  by introducing the angle variables, $\tilde{\theta}_1=\theta_2-\theta_3$, $\tilde{\theta}_2=2\theta_1-\theta_2-\theta_3$,
 and $\tilde{\theta}_3=\theta_1+\theta_2+\theta_3$ for the in-plane angles and  as $\tilde{\phi}_1=\phi_2-\phi_3$, $\tilde{\phi}_2=2\phi_1-\phi_2-\phi_3$,
 and $\tilde{\phi}_3=\phi_1+\phi_2+\phi_3$ for the out-of-plane angles.

We then obtain a linear system decomposed into 6 two-by-two block matrices in which one degree of freedom for rigid motion is paired with an angle variable. For the translation and rotation in $x$ direction, the matrices are given by
\begin{align}
\begin{pmatrix}
C_{D1} & -\frac{\sqrt{3}}{2}k_T \\
-\sqrt{3}k_T & k_F
\end{pmatrix}
\begin{pmatrix}
U_x\\ \dot{\tilde{\theta}}_1
\end{pmatrix}
&=\begin{pmatrix}
\frac{\sqrt{3}}{2}\omega_0K_T\tilde{\theta}_1\\ \kappa\tilde{\theta}_1
\end{pmatrix}
\label{eq:130a},\\
\begin{pmatrix}
C_{R1} & -\frac{\sqrt{3}}{2}(k_T+k_F) \\
-\sqrt{3}(k_T+k_F) & k_F
\end{pmatrix}
\begin{pmatrix}
\Omega_x\\ \dot{\tilde{\phi}}_1
\end{pmatrix}
&=\begin{pmatrix}
\frac{\sqrt{3}}{2}\omega_0(K_F\tilde{\theta}_1-K_T\tilde{\phi}_1)\\ \kappa\tilde{\phi}_1
\end{pmatrix}
\label{eq:130b},
\end{align}
from which we obtain the ordinary differential equations for the linear stability in the form,
\begin{equation}
\frac{d}{dt}
\begin{pmatrix}
\tilde{\theta}_1\\
\tilde{\phi}_1
\end{pmatrix}
=\begin{pmatrix}
A_T & 0\\
A_{TR3} & A_{R3}
\end{pmatrix}\begin{pmatrix}
\tilde{\theta}_1\\
\tilde{\phi}_1
\end{pmatrix}
\label{eq:131a},
\end{equation}
where
$A_T=\Delta_T^{-1}((3/2)|K_Tk_T|\omega_0-|C_{D1}|\kappa)$, $A_{R3}=\Delta_{R3}^{-1}((3/2)|K_T(k_T+k_F)|\omega_0-|C_{R1}|\kappa)$  and $A_{TR3}=(3/2)\Delta_{R3}^{-1}|K_F(k_T+k_F)|\omega_0$, with the determinants $\Delta_T=C_{D1}k_F-(3/2)k_T^2$ and $\Delta_{R3}=C_{R1}k_F-(3/2)(k_T+k_F)^2$. From  (\ref{eq:131a}), we can read off directly the eigenvalues for the  linear stability with the rigid motion in $x$ direction as $A_{T}$ and $A_{R3}$. The eigenvectors are pure rotation around the $x$ axis ($A_{R3}$ eigenvalue)  and   translation along the $x$ axis accompanied by rotation around the $x$ axis ($A_{T}$). 

In a similar manner, we can derive the eigenvalues for linear stability associated with the rigid motion in the $y$ direction. Noting that $C_{D1}=C_{D2}$ and $C_{R1}=C_{R2}$, we have the similar equation to (\ref{eq:131a}), 
\begin{equation}
\frac{d}{dt}
\begin{pmatrix}
\tilde{\theta}_2\\
\tilde{\phi}_2
\end{pmatrix}
=\begin{pmatrix}
A_T & 0\\
A_{TR3} & A_{R3}
\end{pmatrix}\begin{pmatrix}
\tilde{\theta}_2\\
\tilde{\phi}_2
\end{pmatrix}
\label{eq:131b},
\end{equation}
which yields the same eigenvalues $A_{T}$ and $A_{R3}$ as in the linear stability in $x$ direction.

 For the translation and rotation in the $z$ direction, the matrices are now given by
\begin{align}
\begin{pmatrix}
C_{D3} & -k_T \\
-k_T & k_F
\end{pmatrix}
\begin{pmatrix}
U_z\\ \dot{\tilde{\phi}}_3
\end{pmatrix}
&=\begin{pmatrix}
-\omega_0K_T\tilde{\phi}_3\\ \kappa\tilde{\phi}_3
\end{pmatrix},
\label{eq:132a}\\
\begin{pmatrix}
C_{R3} & k_T+k_F \\
3(k_T+k_F) & k_F
\end{pmatrix}
\begin{pmatrix}
\Omega_z\\ \dot{\tilde{\theta}}_3
\end{pmatrix}
&=\begin{pmatrix}
-\omega_0(K_F\tilde{\phi}_3-K_T\tilde{\theta}_3)\\ \kappa\tilde{\theta}_3
\end{pmatrix}
\label{eq:132b},
\end{align}
yielding the equation for the linear stability,

\begin{equation}
\frac{d}{dt}
\begin{pmatrix}
\tilde{\phi}_3\\
\tilde{\theta}_3
\end{pmatrix}
=\begin{pmatrix}
A_{T3} & 0\\
-A_{RT3} & A_{R}
\end{pmatrix}\begin{pmatrix}
\tilde{\phi}_3\\
\tilde{\theta}_3
\end{pmatrix}
\label{eq:133},
\end{equation}
where
$A_{T3}=\Delta_{T3}^{-1}(3|K_Tk_T|\omega_0-|C_{D3}|\kappa)$, $A_{R3}=\Delta_R^{-1}(3|K_T(k_T+k_F)|\omega_0-|C_{R1}|\kappa)$  and $A_{TR3}=3\Delta_R^{-1}|K_F(k_T+k_F)|\omega_0$, with the determinants $\Delta_{T3}=C_{D3}k_F-3k_T^2$ and $\Delta_R=C_{R3}k_F-3(k_T+k_F)^2$. The system in (\ref{eq:133}) provides the eigenvalues of the linear stability associated with the rigid motion in $z$ direction as $A_{R}$ and $A_{T3}$, with corresponding  eigenvectors  of  pure rotation around the $z$ axis and combined translation and rotation along the $z$ axis, respectively. 

In summary, we obtain 6 eigenvalues, $A_{T}$, $A_{T}$, $A_{T3}$, $A_{R}$, $A_{R3}$, $A_{R3}$, each of which is accompanied by a  rigid-motion mode for  the whole cell.

 \subsection{Cell with $N=4$ flagella}
 
 In the case of a cell equipped with $N=4$ flagella, the number of   angle variables exceeds the numbers of degrees of freedom for a rigid-body motion of the cell body.  
We again introduce new angle variables to decompose the square matrix of order $2N+6$ into smaller size systems as $\tilde{\theta}_1=\theta_2-\theta_4$, $\tilde{\theta}_2=\theta_1-\theta_3$, 
$\tilde{\theta}_3=\theta_1+\theta_2+\theta_3$  and
$\tilde{\theta}_4=\theta_1-\theta_2+\theta_3-\theta_4$, with similar combinations for the  out-of-plane angle variables. From these changes of variables, we obtain the 6 matrices associated with the rigid motion and the angle variables, $\tilde{\theta}_i$ and $\tilde{\phi}_i$ for $i=1,2,3$. In turn, we obtain the same form of the linear ordinary differential equations for the linear stability as of  (\ref{eq:131a}), (\ref{eq:131b}) and (\ref{eq:133}), for the $x$, $y$ and $z$ directions, respectively. However, the values of the matrix entries  are  now given by
$A_T=\Delta_T^{-1}(2|K_Tk_T|\omega_0-|C_{D1}|\kappa)$, $A_{R3}=\Delta_{R3}^{-1}(2|K_T(k_T+k_F)|\omega_0-|C_{R1}|\kappa)$ and $A_{TR3}=(2\Delta_{R3}^{-1}|K_F(k_T+k_F)|\omega_0$, with the determinants $\Delta_T=C_{D1}k_F-2k_T^2$ and $\Delta_{R3}=C_{R1}k_F-2(k_T+k_F)^2$ for the instabilities in $x$ and $y$ directions and $A_{T3}=\Delta_{T3}^{-1}(4|K_Tk_T|\omega_0-|C_{D3}|\kappa)$, $A_{R3}=\Delta_R^{-1}(4|K_T(k_T+k_F)|\omega_0-|C_{R1}|\kappa)$ and $A_{TR3}=4\Delta_R^{-1}|K_F(k_T+k_F)|\omega_0$, with the determinants $\Delta_{T3}=C_{D3}k_F-4k_T^2$ and $\Delta_R=C_{R3}k_F-4(k_T+k_F)^2$  for the instability in the $z$ direction.

The remaining two angular degrees of freedom   are diagonalised as 
\begin{equation}
\begin{pmatrix}
k_F & 0 \\
0  & k_F
\end{pmatrix}
\begin{pmatrix}
\dot{\tilde{\theta}}_4 \\
\dot{\tilde{\phi}}_4
\end{pmatrix}
=\begin{pmatrix}
\kappa\tilde{\theta}_4 \\
\kappa\tilde{\phi}_4
\end{pmatrix}
\label{eq:140},
\end{equation}
which leads the two negative eigenvalues for the linear stability, $\kappa/k_F<0$. This, in turn, indicates that the instability can occur only if accompanied by the rigid-body motion of the cell. Here we note that the angle variables, $\tilde{\theta}_4$ and $\tilde{\phi}_4$, do not generate any forces and torque. Inspecting  the definition of $\tilde{\theta}_4$, we see that the force from the angles $\theta_1-\theta_2$ is canceled by that generated by $\theta_3-\theta_4$ and the torque from the part $\theta_1+\theta_3$ cancels   that from $\theta_2+\theta_4$.  

 \subsection{General N case}

The two  simple examples above enable us to now characterise  the instabilities in the general $N$ case. Specifically, we   expect that the instabilities are associated with the rigid-body translation and rotation of the   cell body even when $N\geq 5$. We   introduce the new angle variables as found in the expressions of $\bm{F}_{prop}$ (\ref{eq:90a}) and $\bm{M}_{prop}$ (\ref{eq:90b1})
\begin{align}
\tilde{\theta}_1=\sum_{i=1}^N \theta_i\sin\Theta_i,~~
\tilde{\theta}_2=\sum_{i=1}^N \theta_i\cos\Theta_i,~~
\tilde{\theta}_3=\sum_{i=1}^N \theta_i, \label{eq:150a} \\
\tilde{\phi}_1=\sum_{i=1}^N \phi_i\sin\Theta_i,~~
\tilde{\phi}_2=\sum_{i=1}^N \phi_i\cos\Theta_i,~~
\tilde{\phi}_3=\sum_{i=1}^N \phi_i, \label{eq:150b}.
\end{align}
 
 We first   consider the mode associated with the translation and rotation along the $x$ axis. The resulting $2\times2$ matrices are
 \begin{align}
\begin{pmatrix}
C_{D1} & -k_T \\
-\frac{N}{2}k_T & k_F
\end{pmatrix}
\begin{pmatrix}
U_x\\ \dot{\tilde{\theta}}_1
\end{pmatrix}
&=\begin{pmatrix}
\omega_0K_T\tilde{\theta}_1\\ \kappa\tilde{\theta}_1
\end{pmatrix},
\label{eq:151a}\\
\begin{pmatrix}
C_{R1} & (k_T+k_F) \\
\frac{N}{2}(k_T+k_F) & k_F
\end{pmatrix}
\begin{pmatrix}
\Omega_x\\ \dot{\tilde{\phi}}_1
\end{pmatrix}
&=\begin{pmatrix}
\omega_0(K_F\tilde{\theta}_1-K_T\tilde{\phi}_1)\\ \kappa\tilde{\phi}_1
\end{pmatrix}
\label{eq:151b},
\end{align}
from which we obtain two eigenvalues, 
\begin{align}
A_T&=\Delta_{T}^{-1}(\frac{N}{2}|K_Tk_T|\omega_0-|C_{D1}|\kappa)
\label{eq:151c} ,\\
A_{R3}&=\Delta_{R3}^{-1}(\frac{N}{2}|K_T(k_T+k_F)|\omega_0-|C_{R1}|\kappa)
\label{eq:151d},
\end{align}
where we have introduced the determinants $\Delta_{T}=C_{D1}k_F-(N/2)k_T^2$ and $\Delta_{R3}=C_{R1}k_F-(N/2)(k_T+k_F)^2$ which are positive due to the negative-definiteness of the resistance matrices. The expressions (\ref{eq:151c}) and (\ref{eq:151d}) are similar to the results obtained for $N=3$ and $N=4$.

As expected by symmetry, the modes associated with the translation and rotation along $y$ are similar and  the eigenvalues are  the same as in the $x$ direction, i.e.~$A_T$ and $A_{R3}$.

We then proceed to investigating the modes along the $z$ axis and obtain the $2\times2$ matrices 
 \begin{align}
\begin{pmatrix}
C_{D3} & -k_T \\
-Nk_T & k_F
\end{pmatrix}
\begin{pmatrix}
U_z\\ \dot{\tilde{\phi}}_3
\end{pmatrix}
&=\begin{pmatrix}
\omega_0K_T\tilde{\phi}_3\\ \kappa\tilde{\phi}_3
\end{pmatrix},
\label{eq:152a}\\
\begin{pmatrix}
C_{R3} & (k_T+k_F) \\
N(k_T+k_F) & k_F
\end{pmatrix}
\begin{pmatrix}
\Omega_z\\ \dot{\tilde{\theta}}_3
\end{pmatrix}
&=\begin{pmatrix}
\omega_0(K_F\tilde{\phi}_3-K_T\tilde{\theta}_3)\\ \kappa\tilde{\theta}_3
\end{pmatrix}
\label{eq:152b}.
\end{align}
This system has  eigenvalues 
\begin{align}
A_{T3}&=\Delta_{T3}^{-1}(N|K_Tk_T|\omega_0-|C_{D3}|\kappa),
\label{eq:152c} \\
A_{R}&=\Delta_{R}^{-1}(N|K_T(k_T+k_F)|\omega_0-|C_{R3}|\kappa)
\label{eq:152d},
\end{align}
with  determinants  $\Delta_{T}=C_{D1}k_F-Nk_T^2$ and $\Delta_{R3}=C_{R1}k_F-N(k_T+k_F)^2$ The eigenvalues, (\ref{eq:152c}) and (\ref{eq:152d}), again reproduce the results of $N=3$ and $N=4$.

The eigenvectors obtained so far are associated with  three pure rotational modes and     three combined  translation/rotation modes along the same axis. The remaining degrees of freedom associated with the  other $2N-6$ angular variables do not affect the stability of the cell, which can be summarised to the following statement:
\begin{theorem}\label{T2}
The linear system (\ref{eq:30a}) includes the six modes associated with a rigid-body motion, with  eigenvalues   $A_{T}$, $A_{T}$, $A_{T3}$, $A_{R}$, $A_{R3}$, $A_{R3}$. The remaining  $2N-6$ degrees of freedom all generate identical negative eigenvalues, 
$\kappa/k_F<0$.
\end{theorem}

In order to complete the proof of the above statement, we need to rearrange the angle variables so as to diagonalise  the remaining $2N-6$ degrees of freedom. We first prepare linearly independent $N$ in-plane angle variables, $\tilde{\theta}_i$,  for $i=1, \cdots, N$, in which the angle variables $\tilde{\theta}_1, \tilde{\theta}_2, \tilde{\theta}_3$ defined in (\ref{eq:150a}) are included. When $N\geq 4$, the diagonalisation can then  be achieved if we pick the remaining  angle variables so that they do not generate any effective force and torque. This is possible for an arbitrary $\tilde{\theta}_i$ for $i\geq 4$, since we can add $\tilde{\theta}_1, \tilde{\theta}_2, \tilde{\theta}_3$ without disobeying   the linear independence property for the set of the angle variables. This argument can also be applied to the out-of-plane variables  and the remaining $2N-6$ eigenvalues are found to be all identical and negative,  $\kappa/k_F<0$.

 \subsection{Most unstable mode}

We obtain $2N$ eigenvalues for the linear system, six of which can be positive. In this section, we examine the nature of the most unstable mode, which is the one expected to be relevant in an experiment. 
From  equations (\ref{eq:151c}), (\ref{eq:151d}), (\ref{eq:152c}) and (\ref{eq:152d}), we obtain critical angular velocities, above which the system becomes linearly unstable, as
\begin{align}
\omega_{0T}=\frac{2|C_{D1}|}{N|K_Tk_T|}\kappa
&,~~
\omega_{0T3}=\frac{|C_{D3}|}{N|K_Tk_T|}\kappa
\label{eq:153a},\\
\omega_{0R}=\frac{|C_{R3}|}{N|K_T(k_T+k_F)|}\kappa
&,~~
\omega_{0R3}=\frac{2|C_{R1}|}{N|K_T(k_T+k_F)|}\kappa
\label{eq:153b}.
\end{align}

These values are all positive, indicating that the system is stable in the case where the flagellar filaments pull on the cell body and that the instability can occur only when  flagella push.  
From   equations (\ref{eq:96b}) and  (\ref{eq:96d}), we readily obtain the comparison between the critical values for the in-plane and out-of-plane instabilities as
\begin{equation}
\omega_{0T3} < \omega_{0T}
~~\textrm{  and}~~~
\omega_{0R} < \omega_{0R3}
\label{eq:153c},
\end{equation}
which indicate that the most unstable mode is either the translation towards the $z$ axis (with combined rotation around the same direction) or pure rotation around the $z$ axis. 

Notably, the expressions of the critical values include those of the $N=2$ case, (\ref{eq:81e1}) and (\ref{eq:81e2}). However, due to the symmetry of the system, the eigenvalues are degenerated and the relation (\ref{eq:153c}) becomes equalities  $\omega_{0T3} = \omega_{0T}$ and $\omega_{0R} =\omega_{0R3}$.
Nonetheless, the expressions of $\omega_{0T3}$ and $\omega_{0R}$ in (\ref{eq:153a})-(\ref{eq:153b})
can be obtained when we substitute  the constants   $k_C\mapsto (N/2)k_C$, $k_T\mapsto (N/2)k_T$ and $k_F\mapsto (N/2)k_F$ in (\ref{eq:81e1})-(\ref{eq:81e2}). 
Thus, the rod-like flagellar model examined for  $N=2$ can also be applied in the general $N$ case, and replacing   $c\mapsto(N/2)c$ we obtain the same plots for the ratio of the two critical values, $r$, as in Fig.~\ref{fig_mostunst}(a). This indicates that the increase in the value of $N$ can remove the possibility of $r>1$, and therefore the translation instability would always be  the most unstable mode for pushing flagella. 

We note that   simulation results with $N=4$, which corresponds to a typical number of flagella for \textit{E.~coli} \cite{berg2003}, showed translation in one direction along with   rotation around the same axis \cite{riley2018}, a result consistent with our stability analysis.  We also note that the critical values (\ref{eq:153a}) and (\ref{eq:153b}) do not depend on the value of the torque generated by the flagellar rotation, $K_F$, but they only depend on the force generated by each rotating flagellum. In contrast, the value of $K_F$ appears in the eigenvector of the translation mode and  cells with large $K_F$ undergo rapid rotation.

\section{Small chirality effects}
\label{Sec:chilarity}

The analysis in  the previous sections neglected the diagonal components of the matrix $\mathsf{K}^{(0)}_{TF}$. In this section, we reincorporate the diagonal terms  $k_D$ and solve the resulting   linear stability problem. Since the diagonal components are small  compared to the off-diagonal terms, we may  treat the full problem as a  perturbation  from the results obtained in the previous sections.
 
 As in the previous sections, the  linear stability problem is written  $\mathcal{A}\bm{\Phi}=\bm{b}$ where straightforward calculations now lead the   matrix (see also Appendix A)
\begin{equation}
\mathcal{A}=
\begin{pmatrix}
C_{D1} & 0 & 0 & C_{C1} & 0 & 0 & \bm{C}^{T}_{TF1} & \bm{D}^{T}_{TF1} \\
0  &C_{D2} & 0 &  0&  C_{C2} & 0 & \bm{C}^{T}_{TF2} & \bm{D}^{T}_{TF2} \\
0  & 0& C_{D3} &  0& 0&  C_{C3}  & \bm{D}^{T}_{TF3} & \bm{C}^{T}_{TF3} \\  
C_{C1} & 0 & 0 & C_{R1} & 0 & 0 & \bm{D}^{T}_{TR1} & \bm{C}^{T}_{TR1} \\
0  &C_{C2} & 0 &  0&  C_{R2} & 0 & \bm{D}^{T}_{TR2} & \bm{C}^{T}_{TR2} \\
0  & 0& C_{C3} &  0& 0&  C_{R3}  & \bm{C}^{T}_{TR3} & \bm{D}^{T}_{TR3} \\  
 \bm{C}_{TF1} &  \bm{C}_{TF2}&  \bm{D}_{TF3} &
  \bm{D}_{TR1} & \bm{D}_{TR2} &  \bm{C}_{TR3} &k_F\bm{1}_N & \bm{0}_N \\
   \bm{D}_{TF1} &  \bm{D}_{TF2}&  \bm{C}_{TF3} &
  \bm{C}_{TR1} & \bm{C}_{TR2} &  \bm{D}_{TR3} & \bm{0}_N  &k_F\bm{1}_N 
\end{pmatrix}
\label{Eq:ch02},
\end{equation}
with new  components arising from the chirality given by
\begin{eqnarray}
C_{C1}&=&\sum_{i=1}^N\left(K_T\cos^2\Theta_i +k_D\sin^2\Theta_i\right), \\ 
C_{C2}&=&\sum_{i=1}^N\left(K_T\sin^2\Theta_i +k_D\cos^2\Theta_i\right), \\ 
C_{C3}&=&Nk_D 
\label{Eq:ch03a},
\end{eqnarray}
and new  $N$-dimensional vectors
\begin{eqnarray}
\bm{D}_{TF1}=\bm{D}_{RF1}&=&(k_D\sin\Theta_1,\cdots,k_D\sin\Theta_N)^T, \\ 
\bm{D}_{TF2}=\bm{D}_{RF2}&=&(-k_D\cos\Theta_1,\cdots,-k_D\cos\Theta_N)^T, \\
\bm{D}_{TF3}=-\bm{D}_{RF3}&=&(k_D,\cdots,k_D)^T
\label{Eq:ch03b}.
\end{eqnarray}

With the use of the new angle variables (\ref{eq:150a}) - (\ref{eq:150b}), we may again reduce the linear problem into $4\times4$ blocks associated with the $j$th ($j=1,2,3$) component of the force and torque, while the remaining  degrees of freedom  yield only negative eigenvalues for the stability problem. The explicit form along the $x$ direction can be computed as
\begin{equation}
\begin{pmatrix}
C_{D1} & C_{C1} & C_{TF1} & D_{TF1} \\
C_{C1} & C_{R1} & D_{TR1} & C_{TR1} \\
C_{TF1}  &  D_{TR1} & k_F & 0 \\
D_{TF1} &  C_{TR1}  & 0&k_F 
\end{pmatrix}
 \begin{pmatrix}
 U_1 \\  \Omega_1 \\  \dot{\tilde{\theta}}_1 \\  \dot{\tilde{\phi}}_1
\end{pmatrix}=
 \begin{pmatrix}
 \omega_0K_T \tilde{\theta}_1 \\ \omega_0(K_F\tilde{\theta}_1-K_T\tilde{\phi}_1)\\
 \kappa \tilde{\theta}_1\\ \kappa\tilde{\phi}_1
\end{pmatrix}.
\end{equation}

Inverting the matrix on the left hand side in the previous equation  leads to the following linear ordinary differential equations
\begin{equation}\label{5.9}
\frac{d}{dt}\begin{pmatrix}
\tilde{\theta}_1 \\ \tilde{\phi}_1
\end{pmatrix}=
\begin{pmatrix}
\tilde{A}_T & \tilde{A}_{R3T} \\
\tilde{A}_{TR3} & \tilde{A}_{R3} 
\end{pmatrix}
\begin{pmatrix}
\tilde{\theta}_1 \\ \tilde{\phi}_1
\end{pmatrix},
\end{equation}
for which we can characterise the linear stability as in the previous sections. Using the small parameter $\delta$ to measure the relative magnitude of the chirality effects in the resistance matrix,  $\delta\sim k_D/k_F \sim 10^{-2}$ for the typical parameters of \textit{E.~coli} bacteria, we may expand the matrix in Eq.~\eqref{5.9} at first order in $\delta$ as  $\tilde{A}_T=A_{T} +\delta A'_{T}$, $\tilde{A}_{R3T}=\delta A'_{R3T}$, $\tilde{A}_{TR3}=A_{TR3} +\delta A'_{TR3}$, $\tilde{A}_{R3}=A_{R3} +\delta A'_{R3}$. This results in  similar eigenvalues and eigenvectors for the system with changes of magnitude of order $\delta$. However, as a result of the coupling term $\tilde{A}_{R3T}$,  the pure rotation modes no longer exist and all eigenmodes are now associated with translation in $x$ direction (note that since the changes due to chirality are   small, former rotation  modes are still dominated by rotation and generate small net locomotion).

The analysis along the $y$ and $z$ directions are similar  and we obtain perturbed eigenvectors and eigenvectors from those obtained in the previous sections.  Except for the  6 angle variables, all other modes continue to have stable eigenvalues $\kappa/k_F (<0)$ as shown in the previous section, which can be summarised into the following statement:
\begin{theorem}\label{T3}
The linear stability problem characterised by   matrix (\ref{Eq:ch02}) includes   six modes associated with   rigid-body motion, and the remaining  $2N-6$ degrees of freedom all generate identical negative eigenvalues, 
$\kappa/k_F<0$.
\end{theorem}

The elastohydrodynamic instability  can occur if the flagellar filaments push  the cell body,  accompanied by the rigid motion of the whole cell, with critical angular velocities perturbed from those obtained in the previous section, $\omega_{0T}, \omega_{0T3}, \omega_{0R}, \omega_{0R3}$, since  $\delta$ is   small. 
All   unstable modes now lead to transition  in one direction accompanied by rotation along the same direction. However, there are two types of modes for each direction,  translation-dominated and rotation-dominated modes, the latter of which becomes pure-rotation modes when  chirality effects are neglected.

\bk
\section{Discussion}
\label{sec:conclusion}

In this paper, we     investigated theoretically the elastohydrodynamic stability problem of a model bacterium  with  multiple flagellar filaments  rotated with prescribed frequencies. We assumed that the cell was equipped with $N$ identical flagella connected to a spherical body  surface by a flexible elastic torque spring  and that the flagella are initially arranged in a plane with equal angle intervals as to form a regular $N$-polygon.  We   first formulated the equations of motions of this system  and   showed that this configuration provides an equilibrium state  where the cell body does not move.

 We then   proceeded to consider the linear stability problem  in the case of negligible chirality in the flagellar filaments (active rods).  
When $N=2$, two  modes are obtained (translation and rotation)  which can be unstable when the flagella push on  the cell body provided the magnitude of this pushing force exceeds a critical value (or, for a fixed propulsion magnitude, provided the hook is sufficiently flexible). The translation mode is the more unstable for the typical parameters of real bacteria and  corresponds to the translation in one direction with rotating around the same axis. However, when the flagellar lengths are of the same order as   the cell radius, $L\sim R$, the most unstable mode could be switched to the second mode where the cells rotate in place in the plane of the initial flagellar configuration with no associated translation.  

We then extended our  results   to the general  case of $N$ flagellar filaments,  and  we found that there are always only 6 modes which can be unstable, all of which are associated with  rigid-body motion of the cell. 
The most unstable mode  induces translation towards the  direction perpendicular to the plane in which the $N$ flagella are initially arranged  and   is accompanied by  rotation around the same axis. This analytical result is in agreement with   numerical simulation with $N=4$ helical flagella \cite{riley2018}.

 We finally reincorporated the chirality of the flagellar filaments which had been neglected in the previous sections. Chirality leads to small perturbations of the  eigenvalues and eigenvectors for the linear stability and there are still only 6 possible unstable modes for cells with pusher flagella  associated with rigid motion of the whole cell. The rotation modes  are now accompanied by a small translation of the cell due to chirality-induced coupling.

 The theoretical results in this paper and the presence of  rotation-dominated modes  thus imply  that  multi-flagellated peritrichous bacteria with shorter flagella could fail to swim  efficiently. In contrast, for cells with typical flagellar length $L\sim 10$, the flagella produce a sufficient amount of propulsive thrust
 to lead to an instability to translation. A similar analysis could  also be applied to synthetic particles propelled by bacterial flagella \cite{darnton2004,dileonardo2010} and to the dynamics of an ovum pushed by  multiple spermatozoa \cite{ishimoto2017}.  Note that the linear stability analysis performed in this paper can obviously not fully predict the nonlinear dynamics after the  initial stages of the instability.  
 Furthermore, as shown   in Appendix B, if one considers instead the case of flagella rotated by a constant torque applied in the direction normal to the cell surface, all the possible unstable modes are accompanied by cell translation and   pure-rotation modes disappear. This is in contrast to the case where the  constant torque is applied along the  long axis of the flagellar filament, for which the stability analysis coincides with the fixed-rotation case (see Appendix B). These results   emphasise the  complexity of the multi-flagellated swimming dynamics.

Using typical parameter values for \textit{E.~coli} ($R=1\mu$m, $L=10\mu$m, $N=4$) and  the value of the viscosity for water ($\mu=10^{-3}$Pa$\cdot$s), we can estimate
the critical flagellar rotation rate provided by our theory. The strength of the  torque spring for an \textit{E.~coli} hook has been estimated to be in the range  $\kappa\approx 2.9-8.7\times10^{-21}$Nm using  measured values fo the  hook bending stiffness and length \cite{son2013,riley2018}, and we use the value $\kappa= 5\times10^{-21}$Nm for the following discussions. From the flagellar propulsion force used in Riley {\it et al.} \cite{riley2018} we   have $K_T\approx 7.0\times 10^{-16}$N$\cdot$s. The drag coefficients are estimated as in Sec. \ref{Sec:unst2} with $c'\approx0.13$ and can be used to   obtain an estimation of the critical rotation frequencies for a {\it E.~coli}   cell with $N=4$ flagellar filaments. The critical values for the translation modes are predicted to be $\nu_{0T}\approx 0.36$Hz and  $\nu_{0T3}\approx 0.29$Hz, which are  small compared with those for the pure rotation mode, $\nu_{0R}\approx 1.33$Hz and  $\nu_{0R3}\approx 1.34$Hz. Since the flagellar filaments of real  cells  rotate much faster   ($\nu\approx 100$Hz),  the elastohydroynamic instability obtained in this paper is likely to be relevant  to the locomotion of bacteria.

The model in our paper could be readily extended to the case of a spheroidal cell body if the case where the  flagella are all initially arranged in the equatorial plane of the spheroid, and one would simply need to  change the values of the  drag coefficient $C_D$. When the cell body takes the shape of a prolate spheroid such  as of \textit{E.~coli}, $C_{D1}/C_{D3}>2$ still holds and we obtain the same relation as (\ref{eq:153c}). Other straightforward extensions include the situation in which   the cell body is located near a planar infinite wall, a situation relevant to a sperm-egg cluster that tends to rotate without translation \cite{ishimoto2017}. The predominance of rotation could be rationalised in that case using lubrication theory \cite{lauga2006}, which shows that  drag coefficients $C_{D1}$ and $C_{D3}$ diverge as the spherical cell body approaches the wall, while the value of the rotation drag coefficient $C_{R3}$ remains very close to that in the   bulk. As a result, the rotation mode would become in that case  more unstable than the translation mode.  Further theoretical work would be required to extend to more general situations such as for example a non-spherical cell body, non-symmetric flagellar configurations, or non-identical flagella,  emphasising the rich  diversity of the $N$-flagella problem.

 \vspace{1em}\noindent 
\textbf{Data Accessibility.} This paper has no additional data.\\
\textbf{Authors' Contributions.} KI and EL designed the research and developed the mathematical model. KI analysed the model. KI and EL wrote the paper.\\
\textbf{Competing Interests.} The authors declare that they have no competing interests.\\
\textbf{Funding.} This project has received funding from the European Research Council (ERC) under the European Union's Horizon 2020 research and innovation programme (grant agreement 682754 to EL). KI is supported by MEXT Leading Initiative for Excellent Young Researchers (LEADER), JSPS KAKENHI (Grand Number JP18K13456) and JSPS Overseas Research Fellowship (29-0146).\\
\textbf{Acknowledgements.}  We thank   anonymous referees for helpful comments.\\

\appendix
\section*{Appendix A.  Derivations of  matrices (\ref{eq:75}) and (\ref{Eq:ch02})}
\label{Sec:appA}

In this Appendix, we provide detailed derivations of the matrices (\ref{eq:75}) and (\ref{Eq:ch02}) for the motion of the bacterium with $N$ flagella arranged in a regular polygonal manner.

The $N$ identical in-plane flagella with orientations $\bm{e}^{(i)}=\mathsf{R}(\Theta_i;\bm{e}_z)\cdot\bm{e}_x$ provide a stationary configuration. Considering  small disturbances around the equilibrium, with angles $|\theta_i|, |\phi_i| \ll1$, we   directly compute the matrix entries  noting that one only needs the leading order contributions for the linear stability.

We first consider $\mathsf{K}_C^{(i)}$, which is   computed from (\ref{eq:90a1}),
\begin{eqnarray}
\mathsf{K}_C^{(i)}
=\mathsf{R}_i\cdot\mathsf{K}^{(0)}_C\cdot\mathsf{R}_i^{-1}
\simeq\begin{pmatrix}
K_C\cos^2\Theta_i+k_C\sin^2\Theta_i & (K_C-k_C)\sin\Theta_i\cos\Theta_i & 0\\
(K_C-k_C)\sin\Theta_i\cos\Theta_i & K_C\sin^2\Theta_i+k_C\cos^2\Theta_i  & 0\\
0 & 0 & k_C
\end{pmatrix} 
\label{eq:71a},
\end{eqnarray}
where the symbol $\simeq$ is used here to mean the leading order contribution. The expression for $K_{TT}$ is given by summation of this matrix over the indices $i$.

For the off-diagonal part, we need $\sum_{i=1}^N (\mathsf{K}_C^{(i)}\cdot\mathsf{A}_i'+\mathsf{K}^{(i)}_{TF})$. From   expression (\ref{eq:71a}),  we have
\begin{equation}
\mathsf{K}_C^{(i)}\cdot\mathsf{A}_i'\simeq
\begin{pmatrix}
0 & 0 &  -k_C\sin\Theta_i\\
0 & 0 & k_C\cos\Theta_i\\
k_C\sin\Theta_i & -k_C\cos\Theta_i & 0
\end{pmatrix}
\label{eq:72a},
\end{equation}
and using (\ref{eq:90a1}) we obtain 
\begin{eqnarray}
\mathsf{K}^{(i)}_{TF}\simeq 
\begin{pmatrix}
K_T\cos^2\Theta_i+k_D\sin^2\Theta_i & (K_T-k_D) \sin\Theta_i\cos\Theta_i & -k_T\sin\Theta_i \\
(K_T-k_D)\cos\Theta_i\sin\Theta_i & K_T\sin^2\Theta_i+k_D\cos^2\Theta_i & k_T\cos\Theta_i \\
k_T\sin\Theta_i & -k_T\cos\Theta_i & k_D
\end{pmatrix}
\label{eq:72b}.
\end{eqnarray}

The summations in (\ref{eq:72a}) and (\ref{eq:72b}) give the expression for $\mathsf{K}_{TR}$  and its transpose just follows for $\mathsf{K}_{RT}$. 
 The contributions of the $K_T$ terms can be neglected following the   approximation  showing that  $k_D$ is negligible. For a helical filament, $K_F$ scales as $K_F\sim C_Nb^2L$, and comparing it with the leading-order term we  have the relative magnitude as $K_F/k_F\sim (b/L)^2 \sim 10^{-3}$ using typical numbers for \textit{E.~coli} cells. Thus in the matrix (\ref{eq:75}) the $K_F$ term can be neglected  if we neglect   $k_D$.

For the expression of $\mathsf{K}_{RR}$, we need to calculate 
\begin{equation}
\mathsf{K}_{RR}=C_R\mathsf{1}+\sum_{i=1}^N\left( \mathsf{A}_i^{'T}\cdot\mathsf{K}_C^{(i)}\cdot\mathsf{A}_i'+\mathsf{A}_i^{'T}\cdot\mathsf{K}_{TF}^{(i)}+\mathsf{K}_{FT}^{(i)}\cdot\mathsf{A}'_i+\mathsf{K}_{FF}^{(i)}\right)
\label{eq:73a},
\end{equation}
which is obtained by straightforward calculations as
\begin{equation}
\mathsf{A}_i^{'T}\cdot\mathsf{K}_C^{(i)}\cdot\mathsf{A}_i^{(i)}
\simeq
\begin{pmatrix}
k_C\sin^2\Theta_i & -k_C\sin\Theta_i\cos\Theta_i & 0 \\
-k_C\sin\Theta_i\cos\Theta_i & k_C\cos^2\Theta_i & 0\\
0 & 0& k_C
\end{pmatrix}
\label{eq:73b},
\end{equation}
\begin{equation}
\mathsf{A}_i^{'T}\cdot\mathsf{K}_{TF}^{(i)}
\simeq 
\begin{pmatrix}
k_T\sin^2\Theta_i & -k_T\sin\Theta_i\cos\Theta_i & k_D\sin\Theta_i \\
-k_T\sin\Theta_i\cos\Theta_i & k_T\cos^2\Theta_i & -k_D\sin\Theta_i\\
-k_D\sin\Theta_i & k_D\sin\Theta_i& k_T
\end{pmatrix}
\label{eq:73c},
\end{equation}
\begin{eqnarray}
\mathsf{K}_{FF}^{(i)}
&\simeq&\begin{pmatrix}
K_F\cos^2\Theta_i+k_F\sin^2\Theta_i & (K_F-k_F)\sin\Theta_i\cos\Theta_i & 0\\
(K_F-k_F)\sin\Theta_i\cos\Theta_i & K_F\sin^2\Theta_i+k_F\cos^2\Theta_i  & 0\\
0 & 0 & k_F
\end{pmatrix} 
\label{eq:73d}.
\end{eqnarray}

Using the equalities,
\begin{equation}
\sum_{i=1}^N \sin\Theta_i=\sum_{i=1}^N \cos\Theta_i=\sum_{i=1}^N \sin\Theta_i\cos\Theta_i=0,
\end{equation}
and summing over the index $i$  completes the computations for the matrix entries. 

\section*{Appendix B. Torque-driven motility of $N=2$ flagella}
\label{Sec:appB}

In this Appendix, we briefly consider the bacterial model with $N=2$ flagella in case where the torque, instead of the rotation, is prescribed for each flagellar filament. This will allow us to highlight  the difference of the dynamics from the rotation-given problem. As in the main text, we assume identical flagella and axisymmetric propulsion. Using the same matrix form as (\ref{eq:31c}), the torque-driven motility dynamics can be expressed as
\begin{equation}
\begin{pmatrix}
\mathsf{K}_{TT} & \mathsf{K}_{TR} &\mathsf{K}^{(i)}_{TF}  \\
\mathsf{K}_{RT} & \mathsf{K}_{RR} &\mathsf{K}^{(i)}_{RF}  \\
\mathsf{K}^{(i)}_{FT} & \mathsf{K}^{(i)}_{RT} &\mathsf{K}^{(i)}_{FF}  \\
\end{pmatrix}\begin{pmatrix}
\bm{U} \\
\bm{\Omega} \\
\bm{\omega}^{(i)}
\end{pmatrix}
=\begin{pmatrix}
\bm{0} \\
\bm{0} \\
-\bm{M}^{(i)}_{elast}-\bm{M}^{(i)}_{motor}
\end{pmatrix}
\label{eq:T01}.
\end{equation}

We use the decomposition of the flagellar rotation velocity vector, $\bm{\omega}^{(i)}=\bm{\omega}^{(i)}_t+\bm{\omega}^{(i)}_n$, and the commutative relations (\ref{eq:52c}), to obtain the same form of the force and torque balance equations as  (\ref{eq:42a}),  namely
\begin{equation}
\begin{pmatrix}
\mathsf{K}_{TT} & \mathsf{K}_{TR} &\mathsf{K}^{(i)}_{TF}  \\
\mathsf{K}_{RT} & \mathsf{K}_{RR} &\mathsf{K}^{(i)}_{RF}  
\end{pmatrix}\begin{pmatrix}
\bm{U} \\
\bm{\Omega} \\
\bm{\omega}^{(i)}_n
\end{pmatrix}
=\begin{pmatrix}
-\sum_{i=1}^N \mathsf{K}^{(i)}_{TF}\cdot\bm{\omega}^{(i)}_t \\
-\sum_{i=1}^N \mathsf{K}^{(i)}_{RF}\cdot\bm{\omega}^{(i)}_t 
\end{pmatrix}
\label{eq:T02}.
\end{equation}

The torque balance equation for each flagellum is
\begin{equation}
\mathsf{K}^{(i)}_{FT}\cdot\bm{U}+\mathsf{K}^{(i)}_{FR}\cdot\bm{\Omega}+\mathsf{K}^{(i)}_{FF}\cdot(\bm{\omega}^{(i)}_t+\bm{\omega}^{(i)}_n)=-\bm{M}^{(i)}_{elast}-\bm{M}^{(i)}_{motor}
\label{eq:T03}, 
\end{equation}
which can be rewritten, using the variables in the flagellum-fixed frame, as
\begin{equation}
\mathsf{K}^{(0)}_{FT}\cdot\mathsf{R}^{-1}_i\cdot\bm{U}+\mathsf{R}^{-1}_i\cdot\mathsf{K}^{(i)}_{FR}\cdot\bm{\Omega}+\mathsf{K}^{(0)}_{FF}\cdot(\tilde{\bm{\omega}}^{(i)}_t+\tilde{\bm{\omega}}^{(i)}_n)=-\tilde{\bm{M}}^{(i)}_{elast}-\tilde{\bm{M}}^{(i)}_{motor}
\label{eq:T04}. 
\end{equation}

We next introduce the projection on to $z$ axis as $\mathsf{Q}_0=\mathsf{1}-\mathsf{P}_0$, and apply $\mathsf{P}_0$ and $\mathsf{Q}_0$ from the left side of the equation (\ref{eq:T04}). From the projection onto the $x-y$ plane, we obtain a similar torque balance equation as in the bottom row of the equation  (\ref{eq:42a}), namely
\begin{equation}
\mathsf{P}_0\cdot\mathsf{K}^{(0)}_{FT}\cdot\mathsf{R}^{-1}_i\cdot\bm{U}+\mathsf{P}_0\cdot\mathsf{R}^{-1}_i\cdot\mathsf{K}^{(i)}_{FR}\cdot\bm{\Omega}+\mathsf{K}^{(0)}_{FF}\cdot\tilde{\bm{\omega}}^{(i)}_n=-\tilde{\bm{M}}^{(i)}_{elast}-\mathsf{P}_0\cdot\tilde{\bm{M}}^{(i)}_{motor}
\label{eq:T05},
\end{equation}
noting that the last term of the right-hand side is the only correction from the rotation-given problem.

The projection using $\mathsf{Q}_0$ provides the equations for the tangential flagellar rotation velocity,
\begin{equation}
\mathsf{Q}_0\cdot\mathsf{K}^{(0)}_{FT}\cdot\mathsf{R}^{-1}_i\cdot\bm{U}+\mathsf{Q}_0\cdot\mathsf{R}^{-1}_i\cdot\mathsf{K}^{(i)}_{FR}\cdot\bm{\Omega}+\mathsf{K}^{(0)}_{FF}\cdot\tilde{\bm{\omega}}^{(i)}_t=-\mathsf{Q}_0\cdot\tilde{\bm{M}}^{(i)}_{motor}
\label{eq:T06},
\end{equation}
and we proceed to calculate the detailed expressions for the linear stability analysis around the equilibrium configuration with $N=2$. 

We   need to assume the exact form of the function $\tilde{\bm{M}}^{(i)}_{motor}$, and here we consider two different possibles for the  constant torque: (i) constant torque applied along the flagellar orientation ($\bm{M}^{(i)}_{motor} =M_0\bm{e}^{(i)}$), and (ii) constant torque applied along the normal to the cell surface ($\bm{M}^{(i)}_{motor} =M_0\bm{n}^{(i)}$).

When $|\theta_i|, |\phi_i|\ll1$, the  leading-order value of the right-hand side of (\ref{eq:T06}) is given by $-M_0\bm{e}_z$ in both torque models. The third term on the left-hand side of (\ref{eq:T06}) is simply $\mathsf{K}^{(0)}_{FF}\cdot\tilde{\bm{\omega}}^{(i)}_t=K_F\omega_0\bm{e}_z$, and thus we can neglect $O(|\theta_i|, |\phi_i|)$ contribution in the first two terms of (\ref{eq:T06}) in order to determine the leading-order term of $\omega_0$. With   calculations similar to those  in Appendix A, we obtain the $O(1)$ contribution from the first term as 
$\mathsf{Q}_0\cdot\mathsf{K}^{(0)}_{FT}\cdot\mathsf{R}^{-1}_i\cdot\bm{U}\simeq K_F(\cos\Theta_iU_x+\sin\Theta_i U_y)\bm{e}_z
\label{eq:T07},
$ 
which is however zero as a consequence of the fact that $U_x=\sin\Theta_i=0$ for the linear stability problem with $N=2$. The second term is calculated as
$\mathsf{Q}_0\cdot\mathsf{R}^{-1}_i\cdot\mathsf{K}^{(i)}_{FR}\cdot\bm{\Omega}\simeq K_F(\Omega_x\cos\Theta_i+\Omega_y\sin\Theta_i)\bm{e}_z
$ 
and this is again found to be zero since $\Omega_x$ and $\sin\Theta_i$  are zero for the linear stability problem with $N=2$.

In summary, we obtain the expression for the flagellar rotation rate, $\omega_0$, 
\begin{equation}
\omega_0=\frac{M_0}{|K_F|}
\label{eq:T09}, 
\end{equation}
and  $\omega_0$ becomes positive when $M_0>0$.  Equations (\ref{eq:T02}) and (\ref{eq:T05}) provide therefore  a  set of equations similar to  (\ref{eq:42a})  for the rotation-given problem. We note the presence of the additional term $\mathsf{P}_0\cdot\tilde{\bm{M}}^{(i)}_{motor}$  in   equation  (\ref{eq:T05}). This correction term, however, vanishes for the torque model (i), and thus this torque-driven motility problem coincidences with the rotation-given motility when $N=2$. 

We then consider the torque model (ii) where
the constant torque is applied along the normal to the cell surface. In that case, the correction term contributes   an external bending torque, since   $\mathsf{P}_0\cdot\tilde{\bm{M}}^{(i)}_{motor}\simeq M_0(\theta_i \bm{e}_y - \phi_i \bm{e}_x)$. Proceeding with the linear stability analysis, as in   equations  (\ref{eq:81d01})- (\ref{eq:81d04}) we obtain 4 blocks of $2\times 2$ matrices, using the angle variables $\theta_{+}=\theta_1+\theta_2$, $\theta_{-}=\theta_1-\theta_2$ , $\phi_{+}=\phi_1+\phi_2$ and $\phi_{-}=\phi_1-\phi_2$,
\begin{eqnarray}
\begin{pmatrix}
C'_D & k_T \\
2k_T & k_F 
\end{pmatrix}\begin{pmatrix}
U_z \\
\dot{\phi}_+
\end{pmatrix}
&=&
\begin{pmatrix}
-\omega_0K_T \phi_+ \\
\kappa\phi_+ +M_0\theta_+
\end{pmatrix}\label{eq:T10a1} \\
\begin{pmatrix}
C'_R & k_T+k_F \\
2(k_T+k_F) & k_F 
\end{pmatrix}\begin{pmatrix}
\Omega_z \\
\dot{\theta}_+
\end{pmatrix}
&=&
\begin{pmatrix}
-\omega_0K_T \phi_+  -\omega_0K_F\theta_+\\
\kappa\theta_+ -M_0\phi_+
\end{pmatrix}
\label{eq:T10a2},
\end{eqnarray} 
and
\begin{eqnarray}
\begin{pmatrix}
C'_D & k_T \\
2k_T & k_F 
\end{pmatrix}\begin{pmatrix}
U_y \\
\dot{\theta}_-
\end{pmatrix}
&=&
\begin{pmatrix}
-\omega_0K_T \theta_- \\
\kappa\theta_- -M_0\phi_-
\end{pmatrix} \label{eq:T10b1} \\
\begin{pmatrix}
C'_R & -(k_T+k_F) \\
-2(k_T+k_F) & k_F 
\end{pmatrix}\begin{pmatrix}
\Omega_y \\
\dot{\phi}_-
\end{pmatrix}
&=&
\begin{pmatrix}
\omega_0K_T \phi_-  -\omega_0K_F\theta_-\\
\kappa\phi_- +M_0\theta_-
\end{pmatrix}
\label{eq:T10b2}.
\end{eqnarray} 

Solving   equations (\ref{eq:T10a1})-(\ref{eq:T10a2}) with respect to $\theta_+$ and $\phi_+$, we obtain the linear ordinal differential equations,
\begin{equation}
\frac{d}{dt}\begin{pmatrix}
\theta_+ \\
\phi_+
\end{pmatrix}=\begin{pmatrix}
A_R & A_{RT} \\
A_{TR} & A_T
\end{pmatrix}\begin{pmatrix}
\theta_+\\
\phi_+ 
\end{pmatrix}
\label{eq:T11b}.
\end{equation}
The diagonal components are
\begin{eqnarray}
A_T&=&\Delta_D^{-1}(2\omega_0|K_Tk_T|-\kappa|C'_D|),  \label{eq:T12a}\\
A_R&=&\Delta_R^{-1}(2\omega_0|K_T(k_T+k_F)|-\kappa|C'_R|)  \label{eq:T12b}
\end{eqnarray}
are the same as in the rotation-given problem  (\ref{eq:81d2a})- (\ref{eq:81d2b}). In contrast, 
the off-diagonal components includes the corrections
\begin{eqnarray}
A_{RT}&=&\Delta_R^{-1}(2\omega_0|K_T(k_T+k_F)|+M_0|C'_D|) \label{eq:T13a} \\
A_{TR}&=&\Delta_D^{-1}(-M_0|C'_D|) 
\label{eq:T13b},
\end{eqnarray}
and therefore the stability characteristics could   be different from the rotation-given problem.

When $M_0<0$, the flagellar filaments pull on the cell-body into fluid and we find that the eigenvalues are all negative since $A_T, A_{RT}, A_R<0$ and $A_{TR}>0$. Hence the dynamics is always linearly stable. In contrast, when $M_0>0$, the eigenvalues are still both negative for small values of $M_0$, but become positive when $M_0$  becomes sufficiently large, and there are two critical values of $M_0$ above which instability can occur. Both   unstable modes, however, combine  translation with rotation  and the pure rotation modes disappear (in contrast to the  rotation-given motility problem). The critical flagellar rotation velocities $\omega_0^\ast$ in this case lie between the critical values of the rotation-given problem (\ref{eq:81e1})- (\ref{eq:81e2}) :
\begin{equation}
\min\left\{\omega_{0T},\omega_{0R}\right\}\leq\omega_0^\ast\leq\max\left\{\omega_{0T},\omega_{0R}\right\}
\label{eq:T14}.
\end{equation}  

Note that, by symmetry,   the same eigenvalues follow for the linear ordinal differential equations obeyed by  the variables $\theta_-$ and $\phi_i$ derived from   equations  (\ref{eq:T10b1})-(\ref{eq:T10b2}). 
\bk


\end{document}